\documentclass[11pt,a4paper]{article}
\pdfoutput=1
\usepackage[T1]{fontenc}
\usepackage{jheppub}
\usepackage{rotating}
\usepackage{array}
\usepackage{amsmath}
\usepackage{bm}
\usepackage[normalem]{ulem}
\usepackage{slashed}
\usepackage{booktabs}
\usepackage[pdftex,table]{xcolor}
\usepackage{xspace}
\usepackage{makecell}
\usepackage{url}
\usepackage{todonotes}

\newcolumntype{C}[1]{>{\centering\let\newline\\\arraybackslash\hspace{0pt}}m{#1}}

\newcommand{\abs}[1]{\ensuremath{\lvert#1\rvert}}

\newcommand{\ba}[1]{\ensuremath{\left( #1 \right)}}
\newcommand{\bb}[1]{\ensuremath{\left[ #1 \right]}}
\newcommand{\bc}[1]{\ensuremath{\left\{ #1 \right\}}}

\newcommand{\nocontentsline}[3]{}
\newcommand{\tocless}[2]{\bgroup\let\addcontentsline=\nocontentsline#1{#2}\egroup}

\usepackage{multirow}
\usepackage{caption}
\usepackage{subcaption}

\title{\LARGE When the HL-LHC is blind, LISA is deaf (but not vice versa): 2HDM collider-cosmology synergies}
\author[a,b]{Stefano Moretti,}
\author[a]{André Pousette,}
\author[a,c]{and Carlo Tasillo}

\affiliation[a]{Department of Physics and Astronomy, Uppsala University, Box 516, SE-751 20 Uppsala,
Sweden}

\affiliation[b]{School of Physics and Astronomy, University of Southampton, Highfield, Southampton
SO17 1BJ, United
Kingdom}

\affiliation[c]{Instituto de Física Corpuscular (IFIC), Universitat de València-CSIC, Parc Científic
UV,\\C/ Catedrático José Beltrán 2, E-46980 Paterna, Spain}

\emailAdd{stefano.moretti@cern.ch, andre.pousette.0193@student.uu.se,  carlo.tasillo@ific.uv.es}

\abstract{
Extensions of the Standard Model scalar sector can render the electroweak phase transition
first-order and thereby provide the departure from thermal equilibrium required for electroweak
baryogenesis, while
at the same time sourcing a stochastic gravitational wave (GW) background in the
milli-Hertz range. In this work, we investigate electroweak phase transitions in the CP-conserving
type-I two-Higgs-doublet model (2HDM), focusing on the interplay between collider constraints at the
high-luminosity LHC (HL-LHC) and the projected sensitivity of the Laser Interferometer Space Antenna (LISA). We compute the expected
GW background due to strong first-order electroweak phase transitions and perform extensive
Monte Carlo scans over the collider-viable model parameter space.  We find that GW 
signals within reach of LISA arise almost exclusively in regions of parameter space that necessarily
predict observable deviations at the HL-LHC, in particular through the $H \to ZZ$ decay channel of
the heavy CP-even and neutral Higgs state with $m_H \simeq 180\text{--}250 \, \text{GeV}$. Our results highlight the decisive complementarity between collider and
GW probes and show that the largest parts of the 2HDM parameter space relevant for phase transition signals at LISA can already be tested by HL-LHC. A possible future discovery of 2HDM states at the HL-LHC, however, would not allow conclusive statements about LISA being able to find a GW background due to the amount of parameter tuning required for an observable GW signal.
In order to
evaluate possible caveats of this statement we study the theoretical uncertainties related to the
GW predictions in a two-fold approach using both state-of-the-art tools,
\texttt{BSMPT} and \texttt{TransitionListener}, and also allow for model realizations in which the 
electroweak symmetry is not restored in the high-temperature limit, which are the ones combining
the loudest GW signals with the weakest collider coverage.}


\begin{document}
\maketitle
\flushbottom

\section{Introduction}
\label{sec:introduction}

Despite a decade of high-energy collisions at the Large Hadron Collider (LHC), no
convincing evidence for new particles beyond the Standard Model (SM) has been
found~\cite{ATLAS:2024lda,ATLAS:2024fdw}.  This null result challenges many simple extensions of the
SM, yet it
does not remove the theoretical and cosmological motivations for new physics.
The Higgs boson mass is unstable under quantum corrections, leading to the electroweak
hierarchy problem~\cite{Susskind:1978ms,tHooft:1979rat,Weinberg:1978ym}, and the observed
matter-antimatter asymmetry of the Universe
cannot be explained within the SM alone~\cite{Sakharov:1967dj}.  In particular, the baryon-to-photon
ratio measured in the cosmic microwave background,
$n_B/n_\gamma\simeq6\times10^{-10}$~\cite{Planck:2018vyg},
requires physics beyond the SM because the CP-violating phases in the Cabibbo-Kobayashi-Maskawa
matrix are too small~\cite{Gavela:1993ts,Huet:1994jb} and the electroweak phase transition in
the SM is a smooth crossover~\cite{Kajantie:1996mn,DOnofrio:2015gop}. The
Sakharov conditions for baryogenesis --- baryon number violation, C and CP violation, and
departure from thermal equilibrium --- therefore point to extensions of the scalar sector and
additional sources of CP violation~\cite{Sakharov:1967dj,Morrissey:2012db, vandeVis:2025efm}. These
considerations motivate
searches for physics beyond the SM even in the absence of direct collider signals.
 
One way to realize the required departure from equilibrium is a strong first-order
electroweak phase transition (SFOEWPT), in which bubbles of the broken phase
nucleate and expand in a symmetric plasma.  Extending the scalar sector of the SM can strengthen the
transition, even for the observed Higgs mass around $125 \,
\text{GeV}$~\cite{Morrissey:2012db,Dorsch:2013wja, ATLAS:2012yve, CMS:2012qbp}.  A minimal and
well-studied extension is the two-Higgs doublet model (2HDM),
which introduces an additional scalar doublet and appears in ultraviolet completions such
as supersymmetry, allowing also for a solution of the hierarchy
problem~\cite{Branco:2011iw,Gunion:1984yn}.  In the 2HDM the scalar  sector contains three
electromagnetically neutral ($h, H, A$) and two charged
Higgs states ($H^\pm$), and loop and thermal corrections can induce a potential barrier between the
symmetric
and broken vacua to trigger a SFOEWPT~\cite{Dorsch:2013wja,Basler:2016obg}.  Earlier studies showed
that
regions of the 2HDM parameter space exist where a strong transition is possible, often
favoured by a mass hierarchy with a light SM-like scalar ($h$) and heavier pseudo-scalar states
($A$)~\cite{Dorsch:2013wja,Dorsch:2016nrg,Basler:2016obg,Goncalves:2021egx,Conaci:2026djb}. Such strong
transitions generate a stochastic gravitational wave (GW) background which, for electroweak-scale
transitions, peaks in the milli-Hertz band accessible to space-based interferometers such as the
Laser Interferometer Space Antenna (LISA)~\cite{Caprini:2015zlo, Caprini:2019egz, Caprini:2024hue}.
However, comprehensive scans reveal that such strong transitions
are rare; they typically rely on the existence of a thermally induced barrier with delicate
cancellations with respect to the zero-temperature one-loop
contributions, whereas tree-level or purely radiative barriers are achieved only under additional
model requirements, not necessarily yielding stronger phase transitions in multi-dimensional scalar
potentials~\cite{Bringmann:2026xcx, Coleman:1973jx, Goncalves:2025uwh, Benincasa:2026dhg}.
As we show in section~\ref{subsec:cloudA}, the loudest transitions of the type-I 2HDM are of the
second kind: their barrier is built by the zero-temperature one-loop term rather than thermally
induced. In
practice, most of the
2HDM parameter space yields a crossover or a weak first-order transition, implying that
the associated GW signal is unobservably
small~\cite{Basler:2016obg,Goncalves:2021egx,Goncalves:2023svb}, especially when taking into account
astrophysical GW foregrounds in the mHz band~\cite{Boileau:2022ter,
Caprini:2024hue}.
 
Collider experiments can probe the Higgs potential through searches for additional
scalar states.  The high-luminosity phase of the LHC (HL-LHC)~\cite{Gianotti:2002xx,Azzi:2019yne}
will continue to
test the 2HDM parameter space, in particular through the four-lepton final state $H\to ZZ^{(*)}\to
4\ell$ that
enabled the discovery of the SM Higgs boson~\cite{ATLAS:2012yve,CMS:2012qbp}.  The interplay between
collider and GW observations is therefore particularly valuable: a
GW signal compatible with a SFOEWPT would point to specific
combinations of couplings, masses and mixing angles, while collider searches can
corroborate or exclude these combinations.  Conversely, an absence of deviations at
the HL-LHC would disfavour the possibility of a detectable GW signal
from the 2HDM.
 
Motivated by these considerations, the present work investigates electroweak phase
transitions and collider signatures in the CP-conserving type-I 2HDM. We compute the
finite-temperature effective potential up
to one-loop order, identify regions of parameter space leading to strong first-order
transitions and evaluate the resulting GW spectra. We then analyse the
prospects for observing these scenarios at the HL-LHC via searches for heavy scalar decays. We find
in accordance with previous studies~\cite{Basler:2016obg, Goncalves:2021egx} that two separate
regions of the 2HDM parameter space (referred to as the
\emph{standard-custodial} cloud, $m_A\simeq m_{H^\pm}$, and the \emph{twisted-custodial}
cloud, $m_H\simeq m_{H^\pm}$, see
table~\ref{tab:clouds}) with distinct collider signatures exist in which strong first-order phase
transitions with high signal-to-noise ratios (SNRs) can be found by LISA. Only in the standard-custodial cloud, where $m_A \simeq m_{H^\pm}$, we find
that observable LISA signals can be achieved, corresponding to $\text{SNR} > 10$. A relative amount
of 94.8\,\% of these model realizations can be probed by the HL-LHC, most importantly through the SM
Higgs discovery channel $H \to ZZ$ with subsequent decays to leptons. A complementary check of these
points found by a Monte Carlo (MC) scan maximizing the LISA SNR in the viable 2HDM parameter space using
\texttt{TransitionListener v2}~\cite{Matuszak:2026xsz} (\texttt{TL} in the following) instead of \texttt{BSMPT
v3}~\cite{Basler:2024aaf} finds an even slightly higher testable fraction of 95.9\,\%. Even
loosening the requirement of the restoration of electroweak symmetry (EWSR) at high temperatures,
i.e.~dropping the requirement that $v_1 = v_2 = 0$ at $T \gg 100 \, \text{GeV}$, only decreases
these percentages to 88.7\,\% (84.8\,\%) when using \texttt{BSMPT} (\texttt{TL}) for
the phase transition computations. In the twisted-custodial cloud, where $m_H \simeq m_{H^\pm}$, the SNR stays below the LISA
detection threshold across essentially the entire parameter space, both with and without the requirement of
electroweak symmetry restoration at high temperatures, the sole exception being a handful of points
which cross it only in one of the two codes and only under its more optimistic observation-time
convention.\footnote{Notably, we were \textit{not} able to reproduce the extremely high SNRs of up to
$10^8$ as reported in ref.~\cite{Goncalves:2021egx} in neither of the clouds and with neither
\texttt{BSMPT} nor \texttt{TL}, but instead find maximal SNRs of $141$ ($56$) in the standard-custodial cloud with the \texttt{BSMPT}
(\texttt{TL}) prediction, while the twisted-custodial cloud stays around the LISA
detection threshold. We doubt that SNRs as high as claimed in ref.~\cite{Goncalves:2021egx} can be produced
in any realization of the 2HDM, as we argue in section~\ref{subsec:goncalves}.}
 
These results underline the complementarity of the HL-LHC and LISA in the search for extensions of the
scalar sector of the SM: If the HL-LHC will rule out $H \to ZZ$ decays in the mass interval favoured by
LISA-observable points in the standard-custodial cloud ($m_H \simeq 180\text{--}250 \, \text{GeV})$, this will
dramatically decrease the (Bayesian) likelihood of LISA finding a GW background generated by a strong phase
transition realized in
the 2HDM. The opposite statement, however, does not hold: If, vice versa, the HL-LHC finds hints for
such decays, LISA observations will still be required to make statements about the order and
strength of the electroweak phase transition, as required to make predictions on the viability of
electroweak baryogenesis.
 
This work is structured as follows: In section \ref{sec:GW-2hdm} we introduce the effective
potential of the 2HDM and delineate how we infer the GW signal corresponding to a given model
realization using \texttt{BSMPT} and \texttt{TL}. We then describe the viable 2HDM
parameter space and the methodology used to explore it in a systematic and computationally efficient
way in section~\ref{sec:paramspace}. Our results are described in section~\ref{sec:results} and we
conclude in section~\ref{sec:conclusion}. Appendix~\ref{app:custodial} reviews the custodial-symmetry
structure of the 2HDM that organizes the viable parameter space, appendix~\ref{app:grids} collects the
collider signatures we study, appendix~\ref{app:uncertainties} expands on the uncertainty of the GW
signal predictions, and appendix~\ref{app:corner} provides further plots on the viable parameter space
and the corresponding GW signals.

\begin{table}[t]
  \centering
  \renewcommand{\arraystretch}{1.3}
  \begin{tabular}{l c c c c}
    \toprule
     & $m_A \, / \, \text{GeV}$ & $m_H \, / \, \text{GeV}$ & $m_{H^\pm} \, / \, \text{GeV}$ &
     degeneracy \\
    \midrule
    \textbf{Standard-custodial} & $410\text{--}431$ & $197\text{--}222$ & $420\text{--}437$ & $m_A \simeq m_{H^\pm}$ \\
    \textbf{Twisted-custodial} & $566\text{--}571$ & $177\text{--}189$ & $168\text{--}181$ & $m_H \simeq m_{H^\pm}$ \\
    \bottomrule
  \end{tabular}
  \caption{Favoured mass ranges of the additional scalars in the two
    regions of the type-I 2HDM parameter space leading to observable
    GW signals at LISA. In the standard-custodial cloud we select the
    $\mathrm{SNR} > 10$ region, whereas in the twisted-custodial cloud we use the $\mathrm{SNR} > 1$
    region, since there the LISA SNRs only reach lower values. The quoted ranges are the $16$--$84\,\%$ intervals of the
    Markov Chain MC (MCMC) chains over the EWSR subset. The standard-custodial cloud features a near-degeneracy of the pseudo-scalar $A$ and the charged
    Higgs $H^\pm$, while the twisted-custodial cloud features a near-degeneracy of the heavy
    CP-even scalar $H$ and the charged Higgs $H^\pm$.}
  \label{tab:clouds}
\end{table}

\section{GWs from a phase transition in the 2HDM}
\label{sec:GW-2hdm}

In order to study the phases of the CP-conserving type-I 2HDM and possible transitions between them,
we compute the finite-temperature effective potential in thermal field theory using the
imaginary-time (Matsubara) formalism. We truncate the perturbative expansion at one-loop order and
resum Daisy (ring) diagrams following the Arnold-Espinosa prescription.\footnote{An alternative is
the high-temperature dimensional reduction to a
three-dimensional effective field theory~\cite{Gorda:2018hvi,Croon:2020cgk,Gould:2021oba},
which is gauge invariant and improves the convergence of the perturbative
expansion, but is technically more involved. Non-perturbative lattice
simulations provide the benchmark for the phase
structure~\cite{Andersen:2017ika,Kainulainen:2019kyp},
but are computationally too expensive to be applied to the generic parameter
scans performed in this work. Based on the results found in ref.~\cite{Lewicki:2024xan} we expect
$\mathcal{O}(\%)$-level discrepancies of the LISA-testable 2HDM parameter space between the approach
employed here and the dimensionally reduced theory, i.e.~too small shifts to yield relevant changes
in our inferred collider studies, in which the rates vary smoothly with the scalar masses. The one
place where a per-cent-level mass shift could matter is the $H\to ZZ$ threshold at $2m_Z$, but only
${\sim}5\,\%$ of the LISA-observable points lie within $5\,\%$ of it.} Phase tracing and the
computation of the GW spectrum as well as its observability with LISA are performed using a slightly
adapted version of the code \texttt{BSMPT v3.1.4}~\cite{Basler:2024aaf}, as well as
\texttt{TransitionListener v2.0.1}~\cite{Matuszak:2026xsz}. In the following we will review the
underlying assumptions that went into our computations, pointing out relevant differences between
the two codes. These differences are treated as an estimate of the theoretical uncertainties for the
GW prediction in the following.

\subsection{Computation of the effective potential}
The resummed one-loop effective potential is given by
\begin{align} \label{eq:Veff} 
    V_\text{eff}(\bm{\Phi}, T) = V_\text{tree}(\bm{\Phi}) + V_\text{CW}(\bm{\Phi}) +
    V_\text{CT}(\bm{\Phi}) + V_T(\bm{\Phi}, T) + V_\text{daisy}(\bm{\Phi}, T) \, ,
\end{align}
where $\bm{\Phi} = (\Phi_1, \Phi_2)$ denotes the two Higgs doublets. In the following, we discuss
the individual contributions to the effective potential in the 2HDM. The tree-level scalar potential reads
\begin{align} 
    \label{eq:Vtree}
    V_\text{tree}(\bm{\Phi}) &= m_{11}^2 \abs{\Phi_1}^2 + m_{22}^2 \abs{\Phi_2}^2 - \left[m_{12}^2
    \Phi_1^\dagger\Phi_2 + \text{h.c.}\right] + \frac{\lambda_1}{2}\abs{\Phi_1}^4 +
    \frac{\lambda_2}{2}\abs{\Phi_2}^4 \nonumber\\
    &\quad + \lambda_3\abs{\Phi_1}^2 \abs{\Phi_2}^2
    + \lambda_4 (\Phi_1^\dagger \Phi_2)(\Phi_2^\dagger \Phi_1)
    + \biggl[ \frac{\lambda_5}{2}\left(\Phi_1^\dagger\Phi_2\right)^2 \nonumber\\
    &\qquad + \lambda_6 \left(\Phi_1^\dagger\Phi_1 \right) \left(\Phi_1^\dagger\Phi_2 \right)
    + \lambda_7 \left(\Phi_2^\dagger\Phi_2 \right) \left(\Phi_1^\dagger\Phi_2 \right) + \text{h.c.}
    \biggr] \, ,
\end{align}
with $\Phi_a= (\phi_a^+, \phi_a^0)^\text{T}$ for $a = 1,2$. We restrict ourselves to the
CP-conserving case, such that all parameters are real, and assume a softly broken $\mathbb{Z}_2$
symmetry under $(\Phi_1, \Phi_2)\rightarrow(-\Phi_1, \Phi_2)$, implying $\lambda_{6,7} = 0$. We work
in Landau gauge throughout. The neutral components of both Higgs doublets can be decomposed as
\begin{align}
    \phi_a^0(x) = \frac{1}{\sqrt{2}}\left(\hat{\rho}_a(x) + \mathrm{i}\eta_a(x)\right).
\end{align}
To compute the effective potential we employ the background-field method and split the CP-even
neutral fields as $\hat{\rho}_a(x) = \varphi_a + \rho_a(x)$,
where $\varphi_a$ are homogeneous and static, classical background fields, while $\rho_a$, $\eta_a$,
and $\phi_a^\pm$ denote quantum fluctuations. The effective potential in eq.~\eqref{eq:Veff} is thus
evaluated as a function of $\bm{\Phi}$ which, after the background-field split and upon gauge
fixing, depends only on the classical fields $\varphi_1$ and $\varphi_2$.

At zero temperature, the electroweak vacuum corresponds to the global minimum of the effective
potential, and we define $v_a = \langle \varphi_a \rangle_{T=0}$, with $v_1^2 + v_2^2 =
v_\text{EW}^2 = (246.22 \, \text{GeV})^2$. At tree level, this corresponds to $\langle \Phi_a
\rangle_\text{tree} = \tfrac{1}{\sqrt{2}}(0, v_a)^\text{T}$, which breaks $SU(2)_L \times U(1)_Y$ to
$U(1)_Q$. After electroweak symmetry breaking, three Goldstone modes are eaten by the longitudinal
components of the $W^\pm$ and $Z$ bosons, leaving five physical Higgs states: two CP-even scalars
$h$ and $H$, one CP-odd scalar $A$, and a pair of charged scalars $H^\pm$. The mass eigenstates are
obtained from the interaction eigenstates via
\begin{align}
    \begin{pmatrix} H \\ h \end{pmatrix} = R(\alpha) \begin{pmatrix} \rho_1 \\ \rho_2 \end{pmatrix},
    \qquad
    \begin{pmatrix} G^0 \\ A \end{pmatrix} = R(\beta)\begin{pmatrix} \eta_1 \\ \eta_2 \end{pmatrix},
    \qquad
    \begin{pmatrix} G^\pm \\ H^\pm \end{pmatrix} = R(\beta) \begin{pmatrix} \phi^\pm_1 \\ \phi^\pm_2
    \end{pmatrix},
\end{align}
where $R(\alpha)$ and $R(\beta)$ are orthogonal $2\times2$ rotation matrices and $\tan\beta =
v_2/v_1$. In the chosen Landau gauge, the Goldstone modes $G^0$ and $G^\pm$ remain explicit fields
and are massless at tree level.

The Coleman--Weinberg contribution reads
\begin{align}
    V_\text{CW}(\bm{\Phi}) = \frac{1}{64 \pi^2}\sum_{x}(-1)^{2s_x}(1+2s_x)\,
    m_x^4(\bm{\Phi}) \left(\log \frac{m_x^2(\bm{\Phi})}{\bar{\mu}^2} - k_x\right),
\end{align}
where $x$ runs over all scalar, gauge and fermionic degrees of freedom, $s_x$ denotes the spin, and
$k_x = \tfrac{3}{2}$ ($\tfrac{5}{6}$) for scalars and fermions (gauge bosons). We choose $\bar{\mu}
= v_\text{EW}$ and include the Goldstone modes explicitly, see appendix C.3 of
ref.~\cite{Delaunay:2007wb}. We only consider the Yukawa couplings of the top and bottom quarks and
the tau lepton, which dominate near electroweak symmetry breaking, and fix their coupling structure
to only involve $\varphi_2$, but not $\varphi_1$. This specifies type-I of the 2HDM; alternative
charge assignments which also avoid flavour-changing neutral currents at tree-level by imposing a
$\mathbb{Z}_2$ symmetry in the Yukawa sector define the remaining types and are reviewed in ref.~\cite{Branco:2011iw}. The field-dependent masses of the fermions
and electroweak gauge bosons read
\begin{align}
\label{eq:2HDMmasses}
    m_f^2(\bm{\Phi}) = y_f^2 \varphi_2^2, \qquad
    m_W^2(\bm{\Phi}) = \frac{g_2^2}{4}\left(\varphi_1^2 + \varphi_2^2 \right), \qquad
    m_Z^2(\bm{\Phi}) = \frac{g_1^2 + g_2^2}{4}\left(\varphi_1^2 + \varphi_2^2 \right),
\end{align}
while the photon remains massless at $T=0$ for all field values. The Yukawa and gauge couplings are fixed
such that after EWSB, they match their PDG-recommended value~\cite{ParticleDataGroup:2020ssz}. To
facilitate comparison with collider data, we further impose on-shell-like renormalization conditions
which in turn fix the counterterm parameters $\delta\lambda_i$, $\delta m_{12}^2$, $\delta
m_{11}^2$, and $\delta m_{22}^2$ in $V_\text{CT}(\bm{\Phi})$, following section~3.1 of
ref.~\cite{Basler:2016obg}. That way, the particle masses $m_h$, $m_H$, $m_A$ and $m_{H^\pm}$ as
well as the position of the vevs $v_1$ and $v_2$ are not shifted due to quantum
corrections.

The thermal corrections of the effective potential follow
\begin{align}
    V_T(\bm{\Phi}, T) = \sum_{x}(-1)^{2s_x}(1+2s_x) \frac{T^4}{2\pi^2} J_\pm
    \ba{\frac{m_x^2(\bm{\Phi})}{T^2}},
\end{align}
where
\begin{align}
    J_\pm \ba{\frac{m_x^2(\bm{\Phi})}{T^2}} = \int^\infty_0 \mathrm{d}\boldsymbol{k}\: k^2
    \log\left[1\mp\exp\left(-\sqrt{k^2+\frac{m_x^2(\bm{\Phi})}{T^2}}\right)\right].
\end{align}
Again, $s_x$ denotes the spin of $x$; $J_+$ and $J_-$ are used for bosons and fermions,
respectively. In order to deal with the breakdown of perturbativity for soft bosons with $E\to 0$,
we resum daisy diagrams using the Arnold-Espinosa prescription, i.e.~by adding
\begin{align}
    V_\text{daisy}(\bm{\Phi},T) = -\frac{T}{12\pi}\left[\sum_{b \in {S, V_\text{L}}} n_b\bc{
    \text{Tr}_b \bb{m_b^2(\bm{\Phi}) + \Pi_b(T)}^{3/2}
    - \text{Tr}_b \bb{m_b^2(\bm{\Phi})}^{3/2}
    }
    \right] \, .
\end{align}
Here, the sum runs over only the scalar and longitudinally polarized vector degrees of freedom and
the traces are understood as the sum over the mass eigenvalues of the squared mass matrices with and
without Debye mass corrections in the interaction basis. The hard thermal masses read
\begin{align}
    \Pi_1 &= \frac{12 \lambda_1 + 8 \lambda_3 + 4 \lambda_4 + 3 (3 g_2^2 + g_1^2)}{48} T^2 \,,\\
    \Pi_2 &= \frac{12 \lambda_2 + 8 \lambda_3 + 4 \lambda_4 + 3 (3 g_2^2 + g_1^2) + 12 y_t^2 + 12
    y_b^2}{48} T^2 \, ,\\
    \Pi_{W_\text{L}} &= 2 g_2^2 T^2 \quad \text{and} \quad \Pi_{B_\text{L}} = 2 g_1^2 T^2 \, .
\end{align}

The tree-level potential can hence be parametrized by the set of five parameters
\begin{align}
    \lambda_1,\ \lambda_2,\ \lambda_3,\ \lambda_4,\ \lambda_5,\ m_{12}^2,\ \tan{\beta} \, ,
    \label{eq:coupling_basis}
\end{align}
where $m_{11}^2$ and $m_{22}^2$ have been traded for $v_\text{EW}$ and $\tan\beta$ using the tadpole
conditions. Alternatively, one may work in the mass basis
\begin{align}
    m_h,\ m_H,\ m_A,\ m_{H^\pm},\ \cos(\alpha - \beta),\ m_{12}^2,\ \tan{\beta} \, ,
\end{align}
which we use to infer collider constraints. In practice, we perform this mapping using
\texttt{SPheno v4.0.5}~\cite{Porod:2003um, Porod:2011nf}. The gauge couplings $g_1$ and $g_2$ are
inferred from eq.~\eqref{eq:2HDMmasses} and the observed electroweak gauge boson
masses~\cite{ParticleDataGroup:2020ssz}.

\begin{table}[t]
  \centering
  \begin{tabular}{@{}lrrrrrrr@{}}
    \toprule
    BP & $\lambda_1$ & $\lambda_2$ & $\lambda_3$ & $\lambda_4$ & $\lambda_5$ & $m_{12}^2 /
    \text{GeV}^2$ & $\tan\beta$ \\
    \midrule
    I & 0.276 & 0.264 & 6.09 & $-2.01$ & $-2.26$ & 3291 & 14.8 \\
    II & 0.396 & 0.256 & 7.04 & $-1.81$ & $-1.56$ & 8442 & 14.9 \\
    \bottomrule
  \end{tabular}
  \caption{Input parameters for the two benchmark points used throughout this section. BP~I is an
  ``electroweak symmetry breaking restoring (EWSR)'' point with $m_H \simeq 222\,\text{GeV}$, $m_A \simeq 431\,\text{GeV}$,
  $m_{H^\pm} \simeq 422\,\text{GeV}$ and SNR $\simeq 36$ from the standard-custodial cloud, used in figure~\ref{fig:phases}
  (left panel) and figure~\ref{fig:lisaobservability}. BP~II does not restore electroweak symmetry at $T\gg100\,$GeV, with
  $m_H \simeq 355\,\text{GeV}$, $m_A \simeq 470\,\text{GeV}$, $m_{H^\pm} \simeq 478\,\text{GeV}$ and
  SNR $\simeq 10$ also from the standard-custodial cloud, used in figure~\ref{fig:phases} (right panel). The two points
  belong to the same parameter region (the standard-custodial cloud, $m_A \simeq m_{H^\pm}$) but illustrate the two
  distinct Ultra-Violet (UV) complete scenarios discussed in section~\ref{subsec:Gamma}.}
  \label{tab:benchmark}
\end{table}

\subsection{Bubbles and a macroscopic description of the phase transition}
\label{subsec:Gamma}

\begin{figure}
    \centering
    \includegraphics[width=\linewidth]{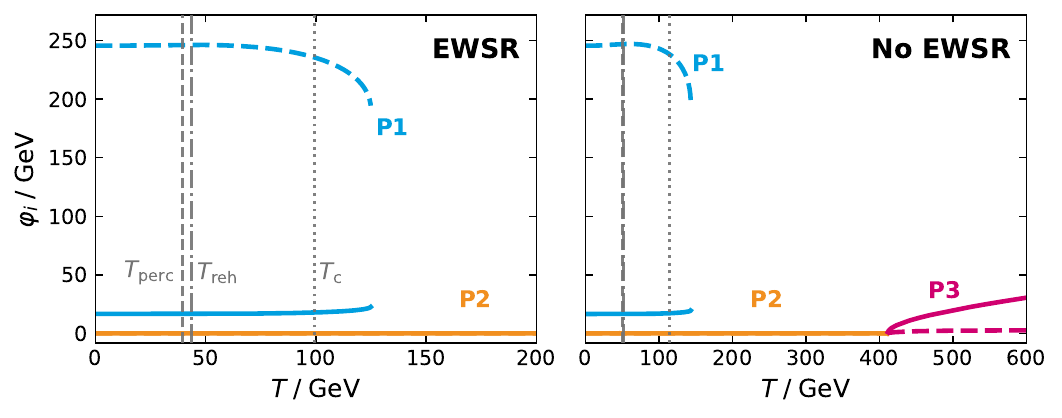}
    \caption{Traced phase structure of the type-I 2HDM as a function of temperature for
    the two benchmark points of table~\ref{tab:benchmark}, obtained with
    \texttt{TL}. Solid and dashed curves show the two background field
    components $\varphi_1$ and $\varphi_2$, with each distinct phase drawn in a separate
    colour. {Left (EWSR, BP~I):} electroweak symmetry is restored at high
    temperature; the broken phase (\textcolor[rgb]{0,0.624,0.875}{cyan}) disappears above
    $T_\text{c}$ and the symmetric phase (\textcolor[rgb]{0.945,0.561,0.122}{orange})
    becomes the global minimum. The vertical lines mark the critical, percolation and
    reheating temperatures of the first-order transition. \emph{Right (No EWSR, BP~II):}
    a high-temperature phase (\textcolor[rgb]{0.816,0,0.435}{magenta}) develops a
    non-vanishing field value via a second-order transition near $T\simeq412\,$GeV, so
    that electroweak symmetry is not restored as $T\to\infty$; the corresponding
    direction is unbounded from below in this limit, which is only admissible for a
    suitable UV completion (see section~\ref{subsec:Gamma}).}
    \label{fig:phases}
\end{figure}

In order to obtain the phase structure of a given realization of the 2HDM, we use
the dependency \texttt{minimum\_tracer} of \texttt{BSMPT} and the module
\texttt{phases} of \texttt{TL}, which were found to give consistent results~\cite{Matuszak:2026xsz}.
The resulting map of phases obtained for BP~I from table~\ref{tab:benchmark} can
be found in the left panel of figure~\ref{fig:phases}, shown for temperatures
from $T = 0$ up to $T = 200 \, \text{GeV}$. For this benchmark point, electroweak
symmetry is restored at high temperatures. At $T \simeq 125 \, \text{GeV}$ a second,
electroweak-breaking minimum appears, which degenerates with the symmetric one at the
critical temperature $T_\text{c} \simeq 99 \, \text{GeV}$ and becomes the global minimum
below, with percolation happening around $T_\text{perc} \simeq 39 \, \text{GeV}$ and a
subsequent reheating to $T_\text{reh} \simeq 44 \, \text{GeV}$. At zero temperature,
the scalar fields in this broken phase lie at $v_1 \simeq 17 \, \text{GeV}$ and $v_2 
\simeq 246 \, \text{GeV}$, ensuring $v_1^2 + v_2^2 = v_\text{EW}^2$. In the right panel of the
same figure, shown up to $T = 600 \, \text{GeV}$, we display an alternative scenario,
corresponding to BP~II from table~\ref{tab:benchmark}. The first-order transition into
the broken phase (which again sits at $v_1 \simeq 17 \, \text{GeV}$, $v_2 \simeq 246 \,
\text{GeV}$ at $T = 0$) is qualitatively similar, with $T_\text{c} \simeq 114 \,
\text{GeV}$, $T_\text{perc} \simeq 50 \, \text{GeV}$ and $T_\text{reh} \simeq 52 \, \text{GeV}$.
However, in this case, the symmetric minimum at $\varphi_1 = \varphi_2 = 0$ shifts to a non-zero
field value for temperatures above $T \simeq 412 \, \text{GeV}$, so that electroweak symmetry is
only restored in an intermediate temperature window  $114 \, \text{GeV} \lesssim T \lesssim 412 \,
\text{GeV}$.
For sufficiently high temperatures, the minimum goes to infinity while the the potential is
unbounded
from below in the limit of $T \to \infty$,
meaning that this scenario can only viable for a given UV completion of the 2HDM, ensuring
boundedness from below and for sufficiently small reheating temperatures after the end of
inflation. We do not generically exclude these cases from our analysis, as they can be
considered features of the 2HDM. In the results presented in section~\ref{sec:results} we
however distinguish the two cases, as the latter requires additional assumptions, whereas
the symmetry-restoring one can be treated as a UV-complete history.

To compute the bubble nucleation rate for transitions between any two phases, we then employ
\begin{align}
    \label{eq:Gamma}
    \Gamma(T)=A(T)\mathrm{e}^{-S_\text{E}(T)} \, ,
\end{align}
where $S_\text{E}(T) = S_3(T) / T$ is the Euclidean bounce action for the transition at a given
temperature and $A(T) \simeq (S_3/(2 \pi T))^{3/2} T^4$. The bounce solution of the scalar fields'
equation of motion is found using the path deformation and a shooting algorithm in both
\texttt{BSMPT} and \texttt{TL}. Using the bubble nucleation rate, we can then compute the
false-vacuum fraction
\begin{align}
    \label{eq:Pfalse}
    P(T) = \mathrm{e}^{-I(T)}\, , \quad \text{where} \quad I(T) \simeq \frac{4\pi
    v_\text{w}^3}{3}\int_{T}^{T_\text{c}} \dfrac{\Gamma (T')\mathrm{d}T'}{T'^4 H(T')} \left
    (\int_T^{T'}\dfrac{d\tilde{T}}{H(\tilde T)}\right)^3 \, .
\end{align}
Here, $v_\text{w}$ is the bubble wall velocity,\footnote{We approximate $v_\text{w}$ using eq.~(3.44)
in ref.~\cite{Basler:2024aaf} in our \texttt{BSMPT} analysis, thus allowing it to vary roughly
between $0.6$ and 1, depending on the transition strength. In the \texttt{TL} analysis, we use the
local thermal equilibrium (LTE) approximation based on the code snippet presented in
ref.~\cite{Ai:2023see}, resorting to the fallback option $v_\text{w}=1$ in case no LTE solution can
be found, see also ref.~\cite{Krajewski:2024zxg}. In appendix~\ref{app:uncertainties} we validate this
approach a-posteriori by computing the wall velocity including out-of-equilibrium pressure terms
using the dedicated code \texttt{WallGo}~\cite{Ekstedt:2024fyq} for a number of benchmark points. In
practice, we find in our \texttt{TL} analysis that the majority of the studied model parameter space
of strong phase transitions hence correspond to $v_\text{w} = 1$. Due to the presence of NLO
friction from gauge boson splitting radiation within the bubble walls~\cite{Gouttenoire:2021kjv}, we
expect terminal, yet relativistic wall velocities in these cases.} $T_\text{c}$ denotes the
critical temperature at which the two phases are degenerate and below which $\Gamma > 0$. We use
eq.~\eqref{eq:Pfalse} to infer the percolation (completion) temperature $T_\text{p}$
($T_\text{compl}$) at which the false vacuum-fraction has dropped to $71\,\%$ ($1\,\%$). Note that
\texttt{TL} uses a generalized expression of the false-vacuum fraction, which does not erroneously
assume an adiabatic expansion of the vacuum-dominated phase and solves $H(T)$ and $P(T)$
self-consistently, see ref.~\cite{Matuszak:2026xsz} for further details.

An often-used reference in order to quantify the strength of a phase transition is based on the vacuum expectation value (VEV)
at the critical temperature~\cite{Moore:1998swa}.
\begin{align}
    \xi_\text{c} = \frac{\sqrt{\langle \phi_1\rangle_{T_\text{c}}^2 + \langle
    \phi_2\rangle_{T_\text{c}}^2}}{T_\text{c}} \, .
\end{align}
This fraction appears in the sphaleron rate; values above $\xi_\text{c} \ge 1$ can be understood as
a proxy for strong phase transitions~\cite{Basler:2024aaf}.

Using the percolation temperature as a reference scale for the point in time when GWs are emitted
(see~\cite{Giese:2020rtr, Giese:2020znk, Jinno:2022mie}), we further compute the transition strength
\begin{align}
    \alpha = \frac{\bar{\theta}_\text{f}(T_\text{perc}) - \bar{\theta}_\text{t}(T_\text{perc})}{3
    (\rho_\text{f}(T_\text{perc}) +p_\text{f}(T_\text{perc})} \overset{\text{BSMPT}}{\simeq} 
    \frac{1}{\rho_\text{rad}} \left[ V_\text{f}
    - V_\text{t}
    - \frac{T}{4}\left(\frac{\partial V_\text{f}}{\partial T} - \frac{\partial V_\text{t}}{\partial
    T}\right)
    \right]_{T=T_\text{p}} \, ,
    \label{eq:alpha_PT}
\end{align}
where
\begin{align}
    \bar{\theta}_i =  \rho_i(T) - \frac{p_i}{c_{\text{s},i}^2} \, , 
\end{align}
is the pseudo-trace~\cite{Giese:2020rtr, Giese:2020znk} of the energy-momentum tensor in the true
($i =\text{t}$) or false ($i =\text{f}$) vacuum, and $c_{\text{s},i}$ is the corresponding speed of
sound. In BSMPT, a bag equation of state is assumed in the computation of $\alpha$, corresponding to
the limit of $c_{\text{s},i}^2 \to 1/3$ in both phases, which breaks down for strong transitions,
see ref.~\cite{Matuszak:2026xsz}.

In both \texttt{BSMPT} and \texttt{TL}, the relevant length scale of the transition is identified
with the mean bubble separation $RH$, which is computed following eqs.~(2.24) in
ref.~\cite{Matuszak:2026xsz} and (3.62) in ref.~\cite{Basler:2024aaf}. The latter again assumes an
adiabatic expansion with $c_{\text{s},\text{f}}^2 = 1/3$, typically slightly overestimating $RH$ for
strong transitions by up to $\mathcal{O}(30\,\%)$.

The redshift of the GW background from its emission to today is computed using the reheating
temperature in both codes. In \texttt{BSMPT}, the approximation 
\begin{align}
    T_\text{reh} \simeq T_\text{perc} \, \min \ba{1, (1 + \alpha)^{1/4}} \, ,
\end{align}
is used for this purpose, whereas \texttt{TL} identifies the temperature, obtained using local
energy conservation,  within the bubbles at the time of percolation with the reheating temperature.
This discrepancy contributes to relative differences in $T_\text{reh}$ of up to $\mathcal{O}(20\,\%)$.
To still allow for a better comparability of the two codes, we use the publicly available
\texttt{TL\_2HDM\_BSMPT.py} model file in the \texttt{TL} analysis, which uses the definition of the
effective degrees of freedom as defined in \texttt{BSMPT}.

\subsection{GW backgrounds}
\label{subsec:GWB}

We can now turn to the computation of the GW signal from the EWPT, given the macroscopic description
of the plasma during the phase transition from the previous subsection. As it is computationally
unfeasible to simulate the nucleation of bubbles on a lattice for each model parameter point of our
scans described in section~\ref{sec:results}, both codes use the semi-analytical GW spectra
recommended by the LISA cosmology working group~\cite{Caprini:2024hue}. These are based on
hydrodynamical simulations and the sound-shell model~\cite{Caprini:2024gyk, Jinno:2022mie, RoperPol:2023dzg}
of a phase transition for a given population of bubbles. The total GW signal is modelled as the sum
of two contributions, stemming from sound waves and turbulence in the plasma after the phase
transition, 
\begin{align}
    h^2\Omega_{\text{GW}}(f) =  h^2\Omega_{\text{sw}}(f) + h^2\Omega_{\text{turb}}(f) \, .
\end{align}
The soundwave contribution follows
\begin{align}
    \label{eq:GWspec}
    h^2 \Omega_\text{sw}(f) = \mathcal{R}h^2 A_{\mathrm{sw}} K^2_\text{sw} \mathcal{Y}_{\mathrm{sw}}
    \ba{RH_*}  S_\text{sw}(f)\,,
\end{align}
where the last term fixes the spectral shape
\begin{align}
  S_\text{sw}(f) &= N {\left( \frac{f}{f_{\text{sw},1}} \right)}^{3} {\left[ 1 + {\left(
  \frac{f}{f_{\text{sw},1}} \right)}^2 \right]}^{-1}
  \left[ 1 + \left( \frac{f}{f_{\text{sw},2}} \right)^4 \right]^{-1},
\end{align}
with normalization
$N=\frac{2\sqrt{2}}{\pi}\left[(1+f_{\text{sw},2}^2/f_{\text{sw},1}^2)
+\sqrt{2}f_{\text{sw},2}/f_{\text{sw},1}\right]$ and two frequency breaks at
\begin{align}\label{eq:fbreaks}
  f_{\text{sw},1} &\simeq  0.2 \ba{\frac{a_*  H_{*}}{RH_*} }\quad \mathrm{and} \quad f_{\text{sw},2}
  \simeq 0.5\, \Delta_{\text{w}}^{-1} \ba{\frac{a_* H_{*}}{RH_*} } \quad \text{with} \quad
  \Delta_\text{w} \equiv \frac{\left| v_{\text{w}} - c_{\text{s}}^\text{bro}
  \right|}{\max(v_{\text{w}}, c_{\text{s}}^\text{bro})} \, .
\end{align}
The Hubble rate at the moment of reheating, red-shifted to today, is given by
\begin{align}\label{eq:Hredshift}
  a_* H_{*} &=  \, 16.5 \, \text{µHz} \,\ba{\frac{T_\text{reh}}{100 \, \text{GeV}}} 
  \ba{\frac{g_{\text{reh}}}{100} }^{1/2}
             \ba{\frac{100}{h_{\text{reh}}}}^{1/3}\,,
\end{align}
where $g_{\text{reh}}$ ($h_{\text{reh}})$ is the effective number of energy (entropy) degrees of
freedom at reheating.

The overall normalization of the spectrum in eq.~\eqref{eq:GWspec} is directly taken from
simulations, with
$A_{\mathrm{sw}} \approx 0.11$, and then further redshifted by a factor of
\begin{align}
  \mathcal{R}h^2 = \ba{\frac{a_\text{reh}}{a_0}}^4 \ba{\frac{H_\text{reh}}{H_0}}^2 h^2
  = \Omega_{\gamma}h^2 \ba{\frac{h_{0}}{h_{\text{reh}}}}^{4/3} \frac{g_\text{reh}}{g_\gamma} \, ,
\end{align}
where $\Omega_\gamma h^2 = 2.473 \cdot 10^{-5}$ is the present  energy
density in radiation~\cite{Planck:2018vyg}, 
$g_\gamma = 2$ and $h_{0} = 3.91$. Finally, we take into account the timescale on which sound waves
contribute to the GW signal by including the factor
$\mathcal{Y}_\text{sw}$. Within \texttt{BSMPT}, this factor is computed following
ref.~\cite{Guo:2020grp}, $\mathcal{Y}_\text{sw} = 1 - \ba{1 + 2 \tau_\text{sh} H_*}^{-1/2}$, where
$\tau_\text{sh} H_* = \min \bb{\frac{2 RH}{\sqrt{3 K_\text{sw}}}, 1}$. In \texttt{TL}, instead, the
lifetime of the soundwave GW sources is estimated as $\mathcal{Y}_\text{sw} = \text{min}\ba{1,
\tau_\text{sh} H_*}$, with $\tau_\text{sh} H_* \simeq 2RH_* / \sqrt{3K_\text{sw}}$,
following~\cite{Caprini:2024hue}. Both variants of the lifetime factor saturate to 1 for long-lived
sources and converge to $\tau_\text{sh} H_*$ for quickly decaying sources $\tau_\text{sh} H_* \ll
1$. They however differ by an $\mathcal{O}(1)$ factor for shells lasting about a Hubble time, see
also figure~1 in ref.~\cite{Guo:2020grp}. The \texttt{TL} description matches the recommendation of
the LISA cosmology working group in ref.~\cite{Caprini:2024hue}, whereas the \texttt{BSMPT}
modelling is based on the assumption of radiation domination, for details see the derivation in
ref.~\cite{Guo:2020grp}. We find that this factor accounts for a large part of the discrepancies
between the two separate analyses.

The turbulence contribution is parametrised by
\begin{align}
    h^2\Omega_\text{turb}(f) = \mathcal{R}h^2 A_{\text{MHD}} K_\text{turb}^2(RH_{\text{sep}})^3
    S_\text{turb}(f)\, .
\end{align}
Here, $A_\text{turb} = 0.255$ and $K_\text{turb} = \epsilon K_\text{sw}$, where $\epsilon = 0.1$.
The spectral shape $S_\text{turb}(f)$ in \texttt{BSMPT} and \texttt{TL} both follow
the lengthy expression provided in eq.~(2.15) in ref.~\cite{Caprini:2024hue}. The precise modelling
of turbulence changes the GW spectrum only minimally due to the smallness of the parameter $\epsilon
= 0.1$ used consistently in both codes. We refer to ref.~\cite{Caprini:2024hue} for an in-depth
description of the used turbulence template and note that the precise modelling does not influence
the conclusions of the present work.

\subsection{Detectability at LISA}
\label{subsec:SNR_LISA}

\begin{figure}
    \centering
    \includegraphics[width=0.7\textwidth]{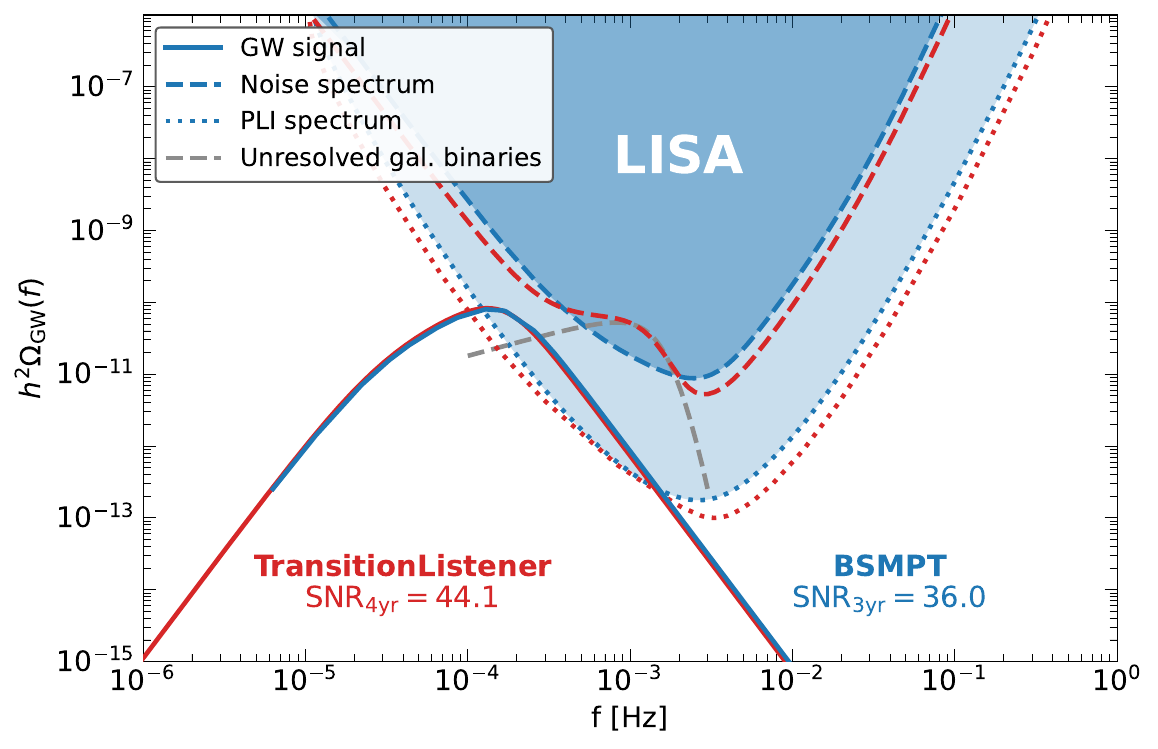}
    \caption{GW spectrum produced by the EWPT described by BP~I, see
    table~\ref{tab:benchmark}. BSMPT (blue) and TransitionListener (red) agree on the detectability
    of the signal at LISA, even though different noise models (dashed curves) are used and the
    signal prediction (solid curves) differs slightly. In particular, the LISA noise budget used in
    TransitionListener assumes the existence of some astrophysical GW foreground due to unresolvable
    galactic binaries (grey dashed curve)~\cite{Robson:2018ifk}. The dotted curves show the
    powerlaw-integrated sensitivity curves of both tools, based on observation periods of $3 \,
    \text{yr}$ ($4 \, \text{yr}$).}
    \label{fig:lisaobservability}
\end{figure}

Given a GW background, both \texttt{BSMPT} and \texttt{TL} compute the expected SNR at LISA
following
\begin{align}
    \text{SNR} = \sqrt{
    \mathcal{T}\int_{f_\text{min}}^{f_\text{max}} \mathrm{d}f \left[
    \dfrac{h^2\Omega_{\text{GW}}(f)}{h^2\Omega_\text{sensitivity}(f)}\right]^2
    }.
    \label{eq:SNR}
\end{align}
Note that \texttt{BSMPT} assumes a data-taking period of $\mathcal{T} = 3 \, \text{yr}$, whereas
\texttt{TL} more optimistically sets $\mathcal{T} = 4 \, \text{yr}$. Further, the LISA noise
modelling in both codes differs slightly, with \texttt{BSMPT} referring to
\cite{LISA_Science_Requirements:2018} and \texttt{TL} using the noise modelling recommended in
ref.~\cite{Breitbach:2018ddu} including an astrophysical foreground of confusion noise due to
unresolved galactic binary mergers, which in turn affects the LISA noise power spectral
density~\cite{Robson:2018ifk}.

In figure~\ref{fig:lisaobservability} we show a comparison of the GW spectrum predicted by
\texttt{BSMPT} as well as \texttt{TL} for the BP~I in table~\ref{tab:benchmark}. The two codes yield
roughly compatible SNRs, based on the previously discussed set of distinct assumptions on details of
the macroscopic description of the phase transition, the lifetime of the soundwave GW sources and
the LISA noise modelling.

Beyond this benchmark-level comparison, \texttt{TL} is integrated into our analysis pipeline as an
independent cross-check of the parameter points making up the standard-custodial cloud and the twisted-custodial cloud (see
section~\ref{sec:results}). For every point retained from the MCMC chains used to construct these
clouds, \texttt{TL} recomputes the SNR at LISA, together with the other phase-transition-related
quantities discussed above (e.g.\ $\alpha$, $\beta/H$, $T_\text{perc}$ and the wall velocity), a
posteriori from the coupling-basis parameters of the corresponding point. This allows us to assess
how sensitive our conclusions are to the choice of code and to the underlying modelling assumptions.

\section{Exploration of the viable 2HDM parameter space}
\label{sec:paramspace} 

In the previous section, we delineated the requirements for a given parameter point in the type-I
2HDM to turn the electroweak phase transition first-order and to emit strong GW signals observable
by LISA. In this section, we turn to the viable regions of the 2HDM parameter space which remain
after imposing a series of observational as well as theoretical constraints. Further, we discuss the
collider implications of these viable points and evaluate their prospects for being probed during
Run~3 of the LHC and at the High-Luminosity LHC (HL-LHC).
We then summarize the different exploration methods we employ in this work in order to study the
collider-sensitivity of the
LISA-testable parts of the viable model parameter space.

\subsection{Theoretical constraints}
First, we consider the theoretical constraints that must be satisfied for any viable realization of
the 2HDM.

\paragraph{Boundedness from below at zero temperature}
A necessary requirement for any realization of the 2HDM to admit a stable (or sufficiently
long-lived metastable) electroweak vacuum with $v_\text{EW} \simeq 246.22\,\text{GeV}$ is that the
tree-level scalar potential is BFB at zero temperature, i.e.\
$\lim_{|\bm{\varphi}|\to\infty} V_\text{tree}(\bm{\varphi}) \neq -\infty$ along any field direction.
For the softly broken $\mathbb{Z}_2$-symmetric potential of eq.~\eqref{eq:Vtree} with
$\lambda_6=\lambda_7=0$, this translates into simple analytic conditions on the quartic
couplings~\cite{Branco:2011iw},
\begin{align}
    \label{eq:BFB}
    \lambda_1 \geq 0\, , \quad \lambda_2 \geq 0\, , \quad \lambda_3 \geq -\sqrt{\lambda_1
    \lambda_2}\, , \quad \lambda_3 + \lambda_4 - \abs{\lambda_5} \geq -\sqrt{\lambda_1 \lambda_2}\,
    ,
\end{align}
which are imposed at the level of the initial parameter scan by \texttt{ScannerS 2.0.1}.

\paragraph{Boundedness from below at finite temperature}
The conditions above only guarantee BFB of the tree-level potential at $T=0$ and do not preclude the
thermally corrected potential from becoming unbounded from below at high temperatures. \texttt{BSMPT
v3.1.4}  allows to perform an additional, separate check of whether $V_\text{eff}(\bm{\Phi},T)$
remains bounded from below in the $T \to \infty$ limit~\cite{Basler:2024aaf}, which amounts to
verifying that the Hessian matrix of the thermal (daisy-resummed) quartic couplings, i.e.\ the
$T^2$-coefficients of the field-dependent thermal masses, remains positive semi-definite. Points
failing this check correspond to potentials that are only metastable for a finite reheating
temperature after inflation and require additional assumptions about the UV completion of the model,
cf.\ the discussion in section~\ref{subsec:Gamma}. By running \texttt{BSMPT} with
\texttt{--checkewsr=on} (see section~3.7.2 in ref.~\cite{Basler:2024aaf} for other options), such
points are retained rather than discarded, which allows us to separately analyse the subset of
points for which electroweak symmetry is restored at high temperatures and the subset for which it
is not.

\paragraph{Vacuum stability at zero temperature}
In addition to being bounded from below, the scalar potential must feature the electroweak vacuum as
its global (or sufficiently long-lived local) minimum at $T=0$. For the CP-conserving 2HDM with
$\lambda_{6,7}=0$ considered here, a tree-level condition ensuring that the electroweak minimum is
the global minimum has been derived in ref.~\cite{Branco:2011iw, Barroso:2013awa} and reads
\begin{align}
    D = m_{12}^2\left(m_{11}^2 - k^2 m_{22}^2 (\tan\beta - k)\right) > 0\, , \qquad k =
    \left(\frac{\lambda_1}{\lambda_2}\right)^{1/4}\, .
\end{align}
This condition is checked by \texttt{ScannerS 2.0.1} for every generated parameter point.

\paragraph{Perturbative unitarity}
Tree-level perturbative unitarity requires that the eigenvalues $M^i_{2\to2}$ of the $2\to2$ scalar
scattering matrix in the high-energy limit satisfy $\abs{M^i_{2\to2}} < 8\pi$~\cite{Branco:2011iw}.
While a violation of this bound does not render the model itself inconsistent, it signals a
breakdown of the perturbative expansion and hence of the validity of perturbative calculations --
such as the one-loop effective potential employed in this work --- for the corresponding parameter
point. \texttt{ScannerS 2.0.1} evaluates the analytic expressions for the eigenvalues of the
scattering matrix of the CP-conserving 2HDM given in ref.~\cite{Branco:2011iw} and discards points
for which perturbative unitarity is violated.

\subsection{Observational constraints}
We now discuss the observational constraints that must be satisfied for any viable realization of
the 2HDM.

\paragraph{Electroweak precision observables}
Electroweak precision observables, such as the $W^\pm$ and $Z$ boson masses, the $Z$ width and the
effective weak mixing angle $\sin^2\theta_W$, are measured to high precision and are summarized in
terms of deviations from their SM predictions via the oblique parameters $S$, $T$ and
$U$~\cite{Branco:2011iw, Peskin:1991sw, Grimus:2007if, Grimus:2008nb, Haller:2018nnx}.
\texttt{ScannerS 2.0.1} computes the 2HDM predictions for $S$, $T$ and $U$ and compares them, via a
$\chi^2$ test, to the global electroweak fit results, excluding parameter points for which $\chi^2$
exceeds the $2\sigma$ threshold.

The tightest of these for the 2HDM is the $T$ parameter, equivalently $\Delta\rho=\rho-1$ with
$\rho\equiv m_W^2/(m_Z^2\cos^2\theta_W)$. It is sensitive to the mass splitting among the extra
scalars, which break the custodial symmetry that protects $\rho=1$. A small $\Delta\rho$ requires
the charged Higgs to be nearly degenerate with one of the neutral non-standard scalars, either
$m_{H^\pm}\simeq m_A$ (the {standard}-custodial case) or
$m_{H^\pm}\simeq m_H$ (the {twisted-custodial} case)~\cite{Gerard:2007kn}, i.e.~the two regions of parameter space
explored in section~\ref{sec:results}. The custodial symmetry of the Higgs sector, the resulting
expressions $\Delta\rho$, and the mass relations are collected in appendix~\ref{app:custodial}.

\paragraph{Flavor constraints}
In the type-I 2HDM with $\lambda_{6,7}=0$, flavour-changing neutral currents are absent at tree
level, and flavour violation beyond the SM only enters through processes mediated by the charged
Higgs boson $H^\pm$. The resulting constraints depend on $m_{H^\pm}$ and $\tan\beta$ and are imposed
by \texttt{ScannerS 2.0.1} in the $(m_{H^\pm}, \tan\beta)$-plane using the same global-fit
results~\cite{Branco:2011iw, Haller:2018nnx} employed for the oblique-parameter constraints.

\paragraph{SM-like Higgs boson}
The properties of the SM-like Higgs boson $h$ must be compatible with the measured properties of the
$125\,\text{GeV}$ Higgs boson discovered at the LHC. This is enforced by \texttt{HiggsSignals
2.6.1}~\cite{Bechtle:2013xfa,Bechtle:2020uwn}, which performs a $\chi^2$ test of the predicted
signal rates and mass of $h$ against the available LHC Higgs measurements, and we require agreement
at the $2\sigma$ level.

\paragraph{Additional Higgs states}
The additional neutral and charged scalars $H$, $A$ and $H^\pm$ must not have been excluded by
direct searches for additional scalar resonances at LEP, the Tevatron and the LHC. This is checked
using \texttt{HiggsBounds 5.9.0}
~\cite{Bechtle:2009ic,Bechtle:2011sb,Bechtle:2013wla,Bechtle:2015pma,Bechtle:2020pkv}, which
compares the predicted production cross sections and branching ratios of all additional Higgs bosons
to the corresponding $95\,\%$~CL exclusion limits, and parameter points excluded at the $2\sigma$
level are discarded.

\subsection{Reach of the HL-LHC}
\label{subsec:hllhc}

Two effects drive the improvement in the expected sensitivity of the HL-LHC compared to current
ATLAS and CMS analyses: the increased luminosity and the higher centre-of-mass energy. The former
can be incorporated through a straightforward statistical rescaling, while the latter modifies the
production cross sections in a process-dependent way due to changes in the parton luminosities. 

To model the effect of the increased integrated luminosity, we rescale the expected upper limits
from ATLAS and CMS by the factor $\sqrt{\mathcal{L}_\text{curr} / \mathcal{L}_\text{HL-LHC}}$, where
$\mathcal{L}_\text{curr}$ is the luminosity of a given $\sqrt{s} = 13 \, \text{TeV}$ ATLAS and CMS
analysis and $\mathcal{L}_\text{HL-LHC}=3~\mathrm{ab}^{-1}$ is the design-sensitivity of the HL-LHC. In
performing this extrapolation, we assume that detector acceptances, selection efficiencies and
systematic uncertainties
remain approximately stable plus signal and background rates scale 
similarly when moving from $13$ to $14\, \text{TeV}$. 

Our study concentrates on the dominant production and decay modes of the additional neutral scalar
and pseudoscalar states, $H$ and $A$, as well as the charged Higgs bosons $H^\pm$, all with properties respecting limits from their null direct searches as implemented in {\tt HiggsBounds},   under the
assumption that the SM-like Higgs boson $h$ remains close to the alignment limit, as enforced by {\tt HiggsSignals}, such that its production and decay 
rates match SM expectations. For each parameter point surviving the EWPT analysis, we compute the
relevant production cross sections and decay branching ratios, then confront them with existing
experimental bounds together with projected sensitivities. 

The collider channels considered in this work are organized by the decaying state, starting with
the heavy CP-even scalar, followed by the CP-odd and the charged Higgs bosons. They are motivated
both by theoretical features of the 2HDM and by the search strategies currently employed at the LHC,
which typically report upper limits on the production rate $\sigma \times \mathrm{BR}$ at
$\sqrt{s}=13~\mathrm{TeV}$ for a given integrated luminosity.

For the heavy CP-even scalar $H$ we consider gluon--gluon fusion (ggF) production followed by the
decays $H \to ZZ$, $H \to W^+W^-$, $H \to \tau^+\tau^-$ and $H \to hh$. The $Z$ boson pair channel,
which once led to the discovery of the SM-like Higgs boson and which will turn out to be the most
constraining one throughout this work, is taken from the CMS search of ref.~\cite{CMS:2026xbb},
covering resonance masses down to $130 \, \text{GeV}$. For $H \to W^+W^-$ we use the CMS limits of
ref.~\cite{CMS:2019bnu} and for $H \to \tau^+\tau^-$ the ATLAS limits of ref.~\cite{ATLAS:2020zms}.
Resonant di-Higgs production is constrained by two ATLAS searches, in the $b\bar b \tau^+\tau^-$ and
in the $b\bar b \gamma\gamma$ final state~\cite{ATLAS:2022xzm, ATLAS:2021ifb}, neither of which is
the more constraining one over the whole mass range relevant here: the $b\bar b \gamma\gamma$ limit
is stronger by a factor $0.6\text{--}0.75$ below $m_H \simeq 330 \, \text{GeV}$, whereas the $b\bar b
\tau^+\tau^-$ limit takes over above $m_H \simeq 350 \, \text{GeV}$ and is up to a factor of $3.7$
stronger at larger masses. We therefore use the more constraining of the two at each resonance mass,
which produces the kink visible in the di-Higgs limit curves of figures~\ref{fig:cloudAgrid}
and~\ref{fig:cloudBgrid}.

For the CP-odd scalar $A$ we use ggF production with the subsequent decay $A \to Zh$ and the
SM-like Higgs boson decaying to a pair of bottom quarks, for which CMS has set $95\,\%$~CL limits on
$\sigma(pp \to A)\times \mathrm{BR}(A\to Zh)\times\mathrm{BR}(h\to b\bar b)$ at $\sqrt{s} =
13$~TeV~\cite{CMS:2019qcx}. As the pseudoscalar is the heaviest of the additional states in both
parameter regions studied below, this is the channel which is sensitive to the upper end of the mass
spectrum.

For the charged Higgs sector, finally, we examine the associated production with a top and a
bottom quark, $gg \to tbH^\pm$, with either the subsequent decay $H^\pm \to \tau^\pm \nu_\tau$,
constrained by ref.~\cite{ATLAS:2024hya}, or $H^\pm \to W^\pm h$ with $h \to b\bar b$, constrained by
ref.~\cite{ATLAS:2024rcu}. To calculate the relevant branching ratios and production cross sections
at $\sqrt{s}=14~\mathrm{TeV}$, \texttt{ScannerS 2.0.1} and \texttt{MadGraph5\_aMC@NLO v3.5.7} were
used, with a custom UFO for the type-I 2HDM.

There are additional production and decay channels of the Higgs companion states that are being pursued
at the LHC (e.g. leading to final states with quarks from the first and second generation), which
offer poor sensitivity in the case of the type-I 2HDM. Similarly for the case of signals
proceeding via $H/A\to Z\gamma$ and $H^\pm\to W^\pm \gamma/Z$ decays. Sensitivity to
processes involving fermionic decays with electrons/positrons is nil for the foreseeable
future at the LHC and upgrades thereof.

EW Precision Observables (EWPOs), primarily stemming from LEP and the SLC (SLAC Linear Collider)
observables but also in part from Tevatron and LHC ones, test custodial symmetry by
constraining deviations in the $\rho$ parameter (the ratio of neutral to charged current
interaction strengths). Custodial symmetry maintains  $\rho\approx1$ at the tree level
and EWPOs confirm this, thereby severely restricting new physics that breaks this symmetry.
However, in a 2HDM, some level of mass degeneracy of the $H^\pm$ state with either or
both the $H$ and $A$ states, as is the case here (see forthcoming figure~\ref{fig:preselection}), ensures that $\rho$ does not depart significantly from
unity. In adopting our experimental constraints from EWPOs we are referring to the
latest datasets from the aforementioned four collider environments. That is, we are
not accounting here for possible projections to the HL-LHC, as the latter will not
significantly improve the precision of the measurement of the $\rho$ parameter (and
other observables \cite{Belvedere:2024wzg}), the main reasons being that such a
measurement requires a clean, single-interaction-per-crossing environment for
elastic scattering and forward/backward proton measurements whereas the HL-LHC is
plagued by high pile-up (a consequence of maximizing the integrated luminosity via
tight beam focusing) and poor high-pseudorapidity acceptance. Furthermore, the HL-LHC
provides a much increased statistics when precision for $\rho$ measurements is actually
limited by systematic uncertainties (like absolute luminosity calibration)
\cite{ATLAS:2022hro}.

\subsection{Exploration strategies}
\label{sec:exploration}

In the present work, we use two main strategies to study the viable parameter space of the 2HDM: A
random scan and a MCMC scan, targeted at finding the regions of parameter space with maximal SNR at
LISA. In the following we describe the technical details of these scans.

\paragraph{Random scan}
Our analysis pipeline begins with a uniform random scan of the type-I 2HDM parameter space, for
which $10\,000$ parameter points are generated with the publicly available code \texttt{ScannerS
2.0.1}~\cite{Muhlleitner:2020wwk}, restricted to
\begin{align}
    \label{eq:param_ranges}
    &m_h = 125.2 \, \text{GeV}\, , \quad
    m_H \in [60,\, 1500] \, \text{GeV}\, , \quad
    m_{A} \in [60,\, 1500] \, \text{GeV}\, , \nonumber \\
    &m_{H^\pm} \in [60,\, 1500] \, \text{GeV}\, , \quad
    c_{HVV} \in [-0.3,\, 0.3]\, , \quad
    m_{12}^2 \in [10^{-3},\, 10^5] \, \text{GeV}^2 \, ,   \\
    &\tan{\beta}\in [0.8,\, 25] \, . \nonumber
\end{align}
where $c_{HVV}$ is the effective gauge coupling of $H$, i.e. the factor that modifies the SM-coupling
between the SM-Higgs and the gauge bosons into the coupling between $H$ and the gauge bosons. In a
$m_h<m_H$ scenario the factor is $c_{HVV}=\cos(\alpha - \beta)$ whereas in the flipped scenario
$m_h>m_H$ the factor becomes $c_{HVV}=\sin(\alpha - \beta)$. For each generated point, \texttt{ScannerS 2.0.1} enforces the
theoretical constraints discussed above -- BFB of the tree-level potential at zero temperature,
vacuum stability and perturbative unitarity --- together with the observational constraints from the
electroweak oblique parameters $S$, $T$, $U$ and from flavour physics, and applies the $2\sigma$
\texttt{HiggsBounds}~\cite{Bechtle:2009ic,Bechtle:2011sb,Bechtle:2013wla,%
Bechtle:2015pma,Bechtle:2020pkv} and \texttt{HiggsSignals}~\cite{Bechtle:2013xfa,%
Bechtle:2020uwn} requirements. Each of the resulting points is then converted
to the coupling-basis parametrization of eq.~\eqref{eq:coupling_basis} by
\texttt{ScannerS 2.0.1} and passed through the publicly available code
\texttt{BSMPT 3.1.4}~\cite{Basler:2024aaf} and its executable \texttt{CalcGW},
which determines the EWPT history of the point together with the associated GW
parameters and the SNR at LISA.

\texttt{BSMPT} is run with \texttt{multistepmode=1}; if this returns a phase tracing warning flag 
\texttt{status\_tracing=no\_glob\_min\_coverage}, indicating that no traced phase is global
over the relevant temperature range, the point is re-run with \texttt{multistepmode=2}. We
further set \texttt{vwall=-2}, a per-point timeout of $1000\,\text{s}$, and keep all remaining
settings at their default values, in particular \texttt{thigh=300}, \texttt{perc\_prbl=0.71},
\texttt{trans\_temp=perc}, \texttt{epsturb=0.1}, \texttt{pnlo\_scaling=1}, \texttt{checknlo=on},
\texttt{checkewsr=on}, \texttt{maxpathintegrations=7}, \texttt{usegsl=true}, \texttt{usenlopt=true}
and \texttt{usemultithreading=false}. Further, the \texttt{--checkewsr=on} setting allows us to
retain and label
points that fail the electroweak symmetry restoration check at $T \to \infty$ rather than discarding
them. The sample therefore splits into two subsets, one in which electroweak symmetry is restored
at high temperature and one in which it is not, and we analyse the two separately. In post-processing we further require that all steps of multi-step transitions are marked with \texttt{check\_compl\_i = success} in order to exclude pathological parts of the 2HDM model parameter space from our analysis.

\paragraph{MC scan}
Because the regions of parameter space giving rise to strong, LISA-detectable GW signals are
sparsely distributed within the seven-dimensional parameter space of eq.~\eqref{eq:param_ranges}, a
purely random scan is inefficient at locating them: the number of samples required to reach a given
sampling density grows exponentially with the number of parameters, better known as the ``curse of
dimensionality''. We therefore complement the initial random scan with a targeted MCMC search,
using the python implementation \texttt{EMCEE 3.1.6}~\cite{Foreman-Mackey:2012any}, chosen
for its ease of integration into our analysis pipeline
and its wide use in the astrophysical literature. The used MCMC algorithm employs an ensemble of
``walkers'' that sample a target probability distribution $p(\bm{x})$ by proposing new points
correlated through the chain of previous samples; after an initial burn-in phase the walkers
converge to sampling $\bm{x}$ with probability $p(\bm{x})$, allowing the ensemble to concentrate on
regions of large $p(\bm{x})$.

In our setup, the sampled probability density is taken to be $p(\bm{x}) \propto
\mathrm{SNR}^2(\bm{x})$, so that the walkers preferentially explore regions of parameter space
yielding large LISA SNRs. Each MCMC run encompasses $140$ walkers, all initialized within a small
hypercube around a single seed point, chosen as the point with the largest SNR among the points
exhibiting a strong first-order electroweak phase transition (SFOEWPT) found in the initial random
scan. The hypercube is populated by drawing random positions in the range $\bb{0.999 x_i, 1.001
x_i}$ for each of the seven model parameter $x_i$ in the coupling basis. Each walker then repeats
the following procedure:
\begin{enumerate}
    \item A parameter point in the coupling basis of eq.~\eqref{eq:coupling_basis} is proposed by
    \texttt{EMCEE}.
    \item \texttt{SPheno v4.0.5} converts this point to the mass basis in order to check if it lies
    within the ranges of eq.~\eqref{eq:param_ranges}, allowing the SM-like Higgs state to have a mass
    within its observationally allowed uncertainty band, $m_h = 125.20\pm0.11$. Then it is passed
    to \texttt{ScannerS 2.0.1}
    and checked against the theoretical and observational constraints discussed above.
    \item If the checks are passed, \texttt{BSMPT} computes the SNR at LISA.
    \item If any of these checks fails, or if \texttt{BSMPT} does not return an SNR due to an
    internal error, the log-likelihood returned to \texttt{EMCEE} is $-\infty$; otherwise it is set
    to $\log(\mathrm{SNR}^2)$. 
\end{enumerate}
For all points surviving the random and MCMC scans, \texttt{ScannerS 2.0.1}, along with\\ \texttt{MadGraph5\_aMC@NLO v3.5.7}, is additionally used to
compute the production cross sections and branching ratios for the collider channels listed above at
$\sqrt{s}=14$~TeV, which form the basis of the HL-LHC sensitivity study in section~\ref{sec:results}.

The MCMC run for the standard-custodial (twisted-custodial) cloud sampled in total approximately
$535{,}000$ ($504{,}000$) points out of which $152{,}000$ ($146{,}000$) points passed the cuts
implemented by \texttt{ScannerS} and were evaluated with \texttt{BSMPT}.
Every point with $\text{SNR}_\texttt{BSMPT} \ge 0.1$, together with those for which
\texttt{BSMPT} did not return an SNR at all, has been processed consecutively with
\texttt{Transition\-Listener v2.0.1}, using the publicly available \texttt{TL\_2HDM\_BSMPT.py} model
file with default numerical accuracies. This allows us to compare the two codes point-by-point along
the chains; the resulting per-point comparisons are presented in appendix~\ref{app:uncertainties}, together
with a brief discussion of the dominant origin of the residual \texttt{BSMPT}--\texttt{TL}
differences. In appendix~\ref{app:uncertainties} we further discuss our use of \texttt{WallGo
v1.1.1}~\cite{Ekstedt:2024fyq, vandeVis:2025plm} which we employed for testing the influence of the
wall velocity on our main findings. A fix to the detonation solver which we contributed in the
course of this work is contained in the subsequent release \texttt{WallGo v1.1.2}.

\section{Results and discussion}
\label{sec:results}

We now present the results of the combined random and MC scans
described in section~\ref{sec:paramspace}. After applying the theoretical
and experimental constraints discussed above, the surviving parameter
points organize into two distinct, well-separated regions of the type-I
2HDM parameter space, which we refer to as the standard-custodial cloud and the
twisted-custodial cloud. Both
clouds feature strong first-order electroweak phase transitions and were
independently identified by the random scan and confirmed and densely
populated by the MCMC search. We first discuss the preselection criterion
used to isolate these two regions (section~\ref{subsec:preselection}) as well as the underlying reason
for the existence of the two clouds (section~\ref{subsec:barrier}), before
turning to the GW and collider phenomenology of the standard-custodial cloud (section~\ref{subsec:cloudA}) and
the twisted-custodial cloud (section~\ref{subsec:cloudB}) individually. Finally,
in sections~\ref{subsec:goncalves} and \ref{subsec:caveats} we comment on the
discrepancy between the SNRs found in this work and the ones found in
ref.~\cite{Goncalves:2021egx} in a similar model setup, and discuss possible caveats
of the strong claim we argue in favour of in this work.

\subsection{Preselection of the standard- and twisted-custodial clouds}
\label{subsec:preselection}

\begin{figure}
    \centering
    \includegraphics[width=0.95\linewidth]{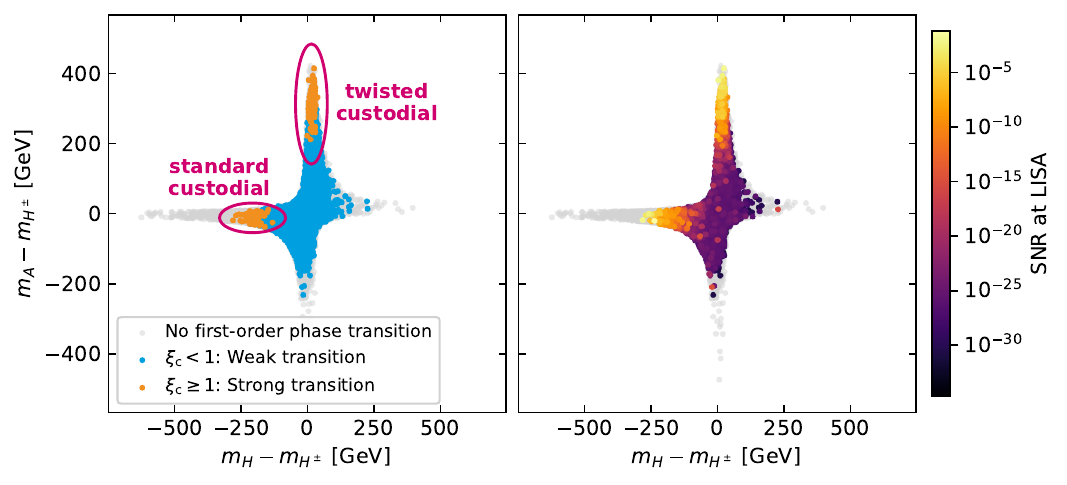}
    \caption{Distribution of the $10{,}000$ points of the random scan in the plane of mass
    differences $m_H - m_{H^\pm}$ and $m_A - m_{H^\pm}$; grey points show points which do not
    feature a first-order phase transition. {Left:} points which feature electroweak
    symmetry restoration at high temperature, and a (strong) first-order phase transition
    are shown in (orange) cyan. Here, we understand strong transitions by requiring $\xi_\text{c} >
    1$. The resulting
    cross-shaped distribution has two arms along the near-degeneracies
    $m_A \simeq m_{H^\pm}$ and $m_H \simeq m_{H^\pm}$, marked by the ellipses labelled
    the standard-custodial cloud the twisted-custodial cloud, cf.~table~\ref{tab:clouds}. {Right:} the same points
    coloured by the LISA SNR.}
    \label{fig:preselection}
\end{figure}

As a first step, in order to get a more intuitive understanding of the six-dimensional parameter
space at hand, we use the transition strength parameter $\xi_\text{c}$
(cf.~section~\ref{subsec:Gamma}) as a preselection criterion, retaining only points
with $\xi_\text{c} > 1$ as candidates for a strong first-order EWPT.
Figure~\ref{fig:preselection} shows the distribution of the mass
differences $m_H - m_{H^\pm}$ and $m_A - m_{H^\pm}$
for the 2HDM realizations produced in the random scan of section~\ref{sec:paramspace},
colour-coded based on whether a first-order phase transition was found (grey),
$0 < \xi_\text{c} \leq 1$ (cyan), or $\xi_\text{c} > 1$ (orange). Of the initial 10,000 points,
$4419$ points survive the boundedness-from-below and completed-transition cuts, and $2955$ of those
feature a first-order phase transition. Of these, $141$ points have a minimum strength of
$\xi_\text{c} \ge 1$.
In the right panel of figure~\ref{fig:preselection}, the same points are coloured by the
SNR at LISA, showing that the points with $\xi_\text{c} > 1$ also yield the highest SNRs,
however, only reaching values far below the detection threshold of $\text{SNR} = 10$.
We conclude that the preselection criterion
$\xi_\text{c} > 1$ is a necessary but not a sufficient condition for a strong first-order
EWPT and for LISA detectability, and an MCMC search is warranted to identify the
regions of parameter space with large SNRs. This is further motivated by taking into account the
small fraction of points with strong first-order EWPTs: Only a percent-level fraction of the
initial random-scan points survive the preselection cut, and none of the preselected points
would be testable at LISA.

The mass-difference distribution exhibits a characteristic
cross-like shape, with two distinct arms extending towards near-degeneracies
$m_A \simeq m_{H^\pm}$ and $m_H \simeq m_{H^\pm}$, respectively. Points with
$\xi_\text{c} > 1$ concentrate almost exclusively along these two arms,
within $|m_A - m_{H^\pm}| \lesssim 30\,\text{GeV}$ and $|m_H - m_{H^\pm}|
\lesssim 30\,\text{GeV}$ of the respective degeneracy lines, corresponding to two physically distinct
ways of realizing a strong phase transition in the type-I 2HDM. The mass ranges
characterizing the two resulting clouds, together with the near-degeneracy pattern
that defines each of them, are summarized in table~\ref{tab:clouds}: the standard-custodial cloud is
characterized by $m_A \simeq m_{H^\pm}$, while the twisted-custodial cloud features $m_H \simeq m_{H^\pm}$.
In appendix~\ref{app:custodial} we motivate the naming of these two clouds and justify the
width of the near-degeneracy bands used to define them, based on the underlying
custodial symmetry of the 2HDM potential and observational limits.

\subsection{Structure of the 2HDM parameter space featuring strong transitions}
\label{subsec:barrier}

\begin{figure}
    \centering
    \includegraphics[width=\textwidth]{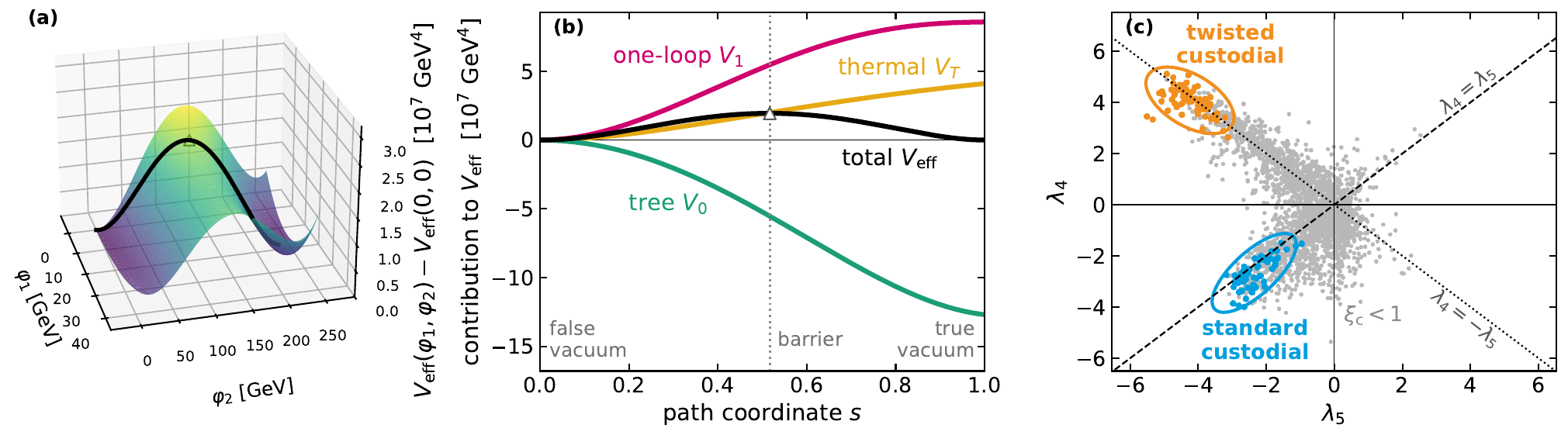}
    \caption{Origin of the electroweak barrier in the type-I 2HDM. (a) The one-loop
    effective potential at the critical temperature $T_\text{c} \simeq 99 \, \text{GeV}$ for the
    benchmark point BP~I of table~\ref{tab:benchmark};
    the black line is the straight path in field space from the symmetric to the broken minimum. (b) The
    potential along this path (black) and its tree-level (green), one-loop $T=0$ (magenta) and
    thermal (yellow) contributions, each relative to the symmetric vacuum. The barrier is built by the one-loop
    Coleman–Weinberg term, while the tree-level opposes it and thermal effects are subleading, in
    agreement with ref.~\cite{Goncalves:2021egx}. (c) The two custodial branches as two lines in the
    $(\lambda_5,\lambda_4)$ plane: the \emph{standard-custodial} cloud along $\lambda_4=\lambda_5$
    ($m_A\simeq m_{H^\pm}$) and the \emph{twisted-custodial} cloud along $\lambda_4=-\lambda_5$
    ($m_H\simeq m_{H^\pm}$); the $141$ points of the
    random scan which restore electroweak symmetry, complete the transition and reach
    $\xi_\text{c}\ge1$ (section~\ref{subsec:preselection}) are highlighted. Both clouds require a large
    $-\lambda_5$, i.e.~a heavy non-decoupling pseudoscalar~$A$ whose mass is induced by electroweak
    symmetry breaking.}
    \label{fig:barrier}
\end{figure}

We now come to the underlying reason for the existence of the two clouds identified in
the previous section. As a result, we will be able to get a more intuitive understanding
of the six-dimensional parameter space and understand the size of the quartic
couplings and mass ranges which LISA will be able to probe. 

We find that the potential barrier separating the symmetric and broken minima is
generated dominantly by the one-loop (Coleman--Weinberg) part of the effective
potential at zero temperature, $V_1=V_\mathrm{CW}+V_\mathrm{CT}$, while the tree-level
potential typically \emph{opposes} the barrier and the thermal corrections play a
subleading role. This is shown for our random scan in figure~\ref{fig:barrier} in
spirit of the analysis in ref.~\cite{Goncalves:2021egx}:
for $\xi_\text{c}>1$ the barrier is increasingly one-loop dominated. In particular it
is \emph{not} a thermally induced (cubic) barrier: the daisy resummation cancels the
$\phi^3 T$ term of the scalars and longitudinal gauge bosons, and $\lambda_5$, argued below
to be the decisive parameter, does not even enter the Debye masses, so the surviving thermal contribution
is subdominant and largely SM-like.

Because the dominant contribution is the zero-temperature one-loop term, the strength of the
transition is controlled by a zero-temperature quantity: the \emph{upliftment} of the true vacuum,
i.e.~the zero-temperature free-energy gap $\Delta \mathcal{F}_0$ between the symmetric and the broken
minimum~\cite{Dorsch:2017nza,Harman:2015gif,Chung:2012vg}. The larger this gap, the more vacuum energy
is released and the stronger the transition. In the alignment limit $\Delta \mathcal{F}_0$ is governed by the \emph{non-decoupling} parts of the
heavy-scalar masses, $m_X^2-M^2=-\lambda_X v_\text{EW}^2$, where
$M^2\equiv m_{12}^2/(\sin\beta\cos\beta)$ is the soft (decoupling) scale. The non-decoupling part is
thus the piece of each mass sourced by electroweak symmetry breaking rather than by $M$. The pseudoscalar combination
$m_A^2-M^2=-\lambda_5 v_\text{EW}^2$ dominates: a large negative $\lambda_5$ (a heavy, strongly
non-decoupling $A$) produces a large one-loop barrier and hence a strong
transition, cf.~ref.~\cite{Dorsch:2013wja}. Both clouds require such an $A$, with
$\lambda_5\simeq-2.1$ ($m_A\simeq420\,$GeV) in the standard-custodial cloud and $\lambda_5\simeq-5.0$
($m_A\simeq570\,$GeV) in the twisted-custodial one.

The two clouds differ in how the custodial degeneracy is realized, i.e.\ in $\lambda_4$
(figure~\ref{fig:barrier}). In the standard-custodial cloud $\lambda_4\simeq\lambda_5\simeq-2.3$, so the
charged Higgs is dragged up together with $A$ ($m_A\simeq m_{H^\pm}\simeq420$--$430\,$GeV) and both feed
the upliftment. In the twisted-custodial cloud $\lambda_4\simeq-\lambda_5\simeq+4.7$, so $H^\pm$ and the
CP-even $H$ stay light near the soft scale ($m_H\simeq m_{H^\pm}\simeq170$--$200\,$GeV) and the
upliftment is provided by the even heavier $A$ alone.

We now discuss the remaining portal couplings. The portal combination
$\lambda_{345}=\lambda_3+\lambda_4+\lambda_5$
sets the coupling of the SM-like Higgs to the heavy scalars and, in the alignment limit, is tied to the
heavy CP-even mass~\cite{Branco:2011iw}, and follows from the pseudoscalar, charged and CP-even
mass relations of ref.~\cite[eqs.~(10)--(13)]{Gunion:2002zf},
\begin{equation}
    m_H^2 - M^2 = \sin^2\!\beta\,\cos^2\!\beta\,\left(\lambda_1+\lambda_2-2\lambda_{345}\right)v_\text{EW}^2 \, ,
    \label{eq:mH_align}
\end{equation}
whose prefactor is set by $\tan\beta$, so the two clouds behave differently. In the twisted-custodial
cloud ($\tan\beta\simeq3.9$) the prefactor is sizeable, and the light, near-degenerate $H$
($m_H\simeq m_{H^\pm}$) fixes the portal tightly to a small value, $\lambda_{345}\simeq0.2$, meaning
that the extra scalars nearly decouple from the $125\,$GeV Higgs. In the standard-custodial cloud, instead,
the large
$\tan\beta\simeq13.5$ suppresses the prefactor (through $\cos^2\!\beta\simeq0.005$), so $m_H\simeq M$
holds irrespective of $\lambda_{345}$ and the portal coupling $\lambda_{345}$ is no longer set by the
mass spectrum. Instead it is
controlled by boundedness-from-below: with $\lambda_4\simeq\lambda_5$ this requires
$\lambda_{345}=\lambda_3+2\lambda_5>-\sqrt{\lambda_1\lambda_2}$, cf.~eq.~\ref{eq:BFB},
so the large negative $\lambda_5$
demanded by a strong transition must be offset by a comparably large $\lambda_3$, leaving
$\lambda_{345}$ small ($\simeq0.2$--$0.4$ across the bulk of the preselected cloud) and rising to
$\simeq1$--$2$ only for the loudest, most strongly supercooled points. The Higgs quartic
$\lambda_2\simeq0.26$ is fixed by $m_h=125\,$GeV, and $\lambda_1$ enters only through the small VEV
$v_1=v_\text{EW}\cos\beta$, so its effect on the effective potential is subdominant:
$\lambda_1 v_1^4 \ll \lambda_2 v_2^4$. In practice, in the following MC analysis, we therefore
find $\lambda_1\simeq0.3$--$1.5$ ($\lambda_1\simeq3.7$--$5.1$) in the standard-custodial (twisted-custodial)
cloud, each one with a large spread of allowed values, i.e.~little impact on the transition strength.
In either cloud the transition strength is hence governed by the single ``barrier'' coupling
$-\lambda_5$. These relations, inferred here from the barrier structure and the alignment
limit, are confirmed empirically by the MC scans of the following two subsections
and can be verified in the corner plots in appendix~\ref{app:corner}.

In summary, we can conclude that two arms follow from the interplay
between a strong transition and electroweak precision data.
A strong transition requires large quartic couplings, hence large mass splittings among the additional
scalars, which break the custodial symmetry and drive up $\Delta\rho$; precision data force it back
down. The tension is resolved only on one of the two custodial degeneracies,
$m_{H^\pm}\simeq m_A$ or $m_{H^\pm}\simeq m_H$~\cite{Gerard:2007kn}, which are
precisely the two arms of figure~\ref{fig:preselection}. We expand on the custodial limits in appendix~\ref{app:custodial}.

\subsection{GWs and collider-testability in the standard-custodial cloud}
\label{subsec:cloudA}

The standard-custodial cloud sits on the custodial limit $\lambda_4\simeq\lambda_5$,
in which the pseudoscalar and the charged Higgs are nearly degenerate,
$m_A\simeq m_{H^\pm}$. The chains are seeded at 140 slightly shifted points around the highest-$\xi_\text{c}$ point of the
standard-custodial cloud (see section~\ref{sec:exploration}) and run for $3{,}821$ steps each,
sampling $534{,}940$ points in total. Of these, $152{,}307$ pass the cuts of
section~\ref{sec:exploration}, about $141{,}600$ of them distinct after de-duplicating repeated
chain entries, and they populate the standard-custodial mass window of table~\ref{tab:clouds}. Throughout the cloud the heavy CP-even scalar remains above the observed
SM-like Higgs, $m_H>m_h$; the $m_H<m_h$ region that showed up in the initial random scan and is allowed by
our prior ranges in eq.~\eqref{eq:param_ranges} is never populated by the MCMC.

\begin{figure}
    \centering
    \includegraphics[width=0.85\linewidth]{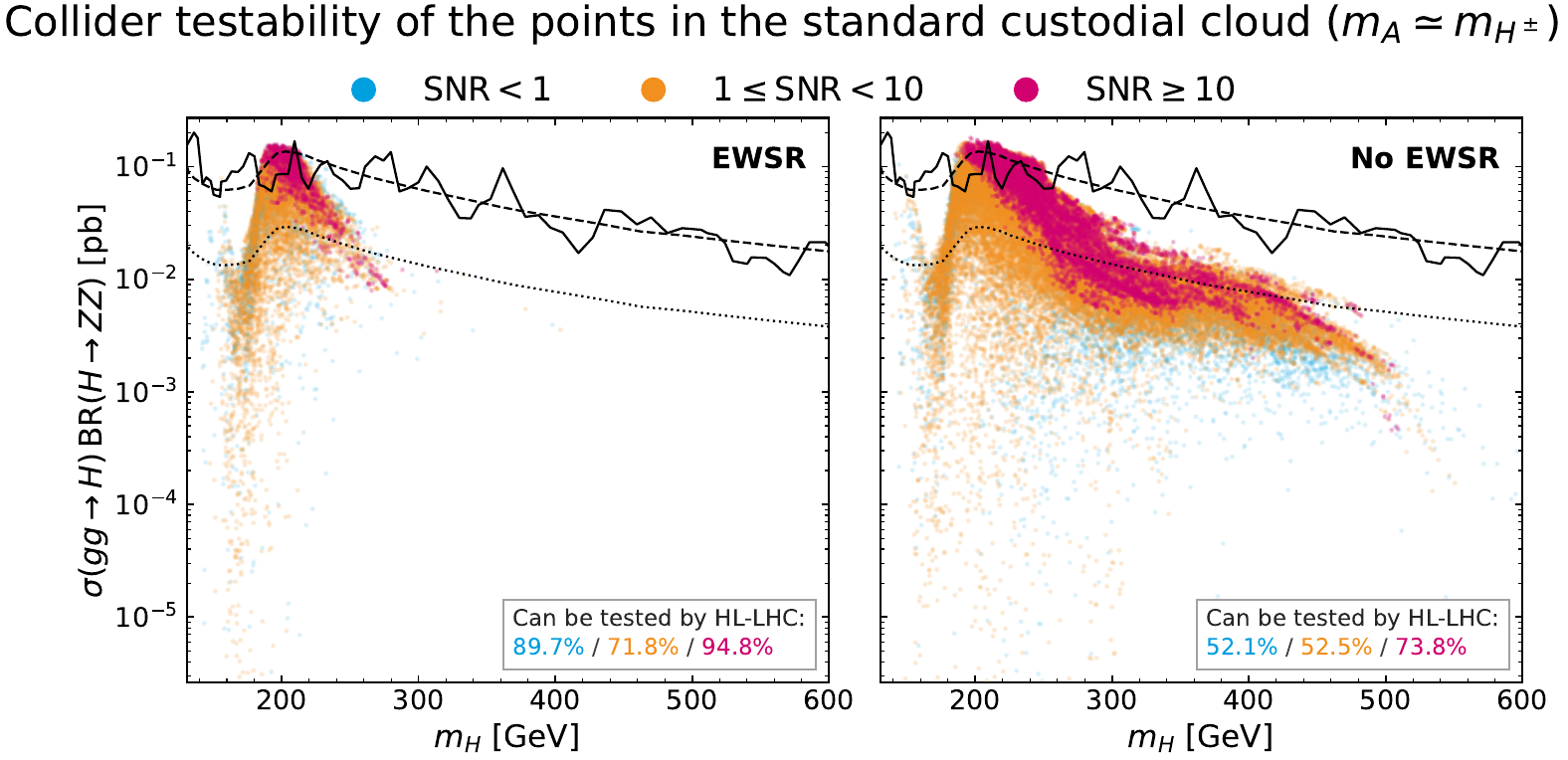}
    \caption{Predicted cross section $\sigma(gg \to H)\, \text{BR}(H \to ZZ)$ at
    $\sqrt{s}=14\,\text{TeV}$ as a function of $m_H$ for the points of the standard-custodial cloud, for the subset with
    electroweak symmetry restoration at high temperature (EWSR, left) and the non-restoring subset (No EWSR, right).
    The solid and dashed curves show the current observed and expected upper limits on the effective
    $H \to ZZ$ cross section, respectively, and the
    dotted curve shows the projected HL-LHC sensitivity. 94.8\,\% (73.8\,\%) of the shown LISA-observable model
    realizations can be tested by the HL-LHC, when requiring that the electroweak symmetry is (not) restored at high
    temperatures.}
    \label{fig:cloudA_zz}
\end{figure}

In our analysis, we keep separate the two thermal histories introduced in section~\ref{subsec:Gamma},
which form the two columns of figures~\ref{fig:cloudA_zz} and~\ref{fig:cloudAgrid}. In the
{EWSR} subset electroweak symmetry is restored at high temperature and the transition proceeds
from the symmetric origin in the textbook way (benchmark BP~I of table~\ref{tab:benchmark}); in the
{No EWSR} subset the high-$T$ potential is unbounded from below along the transition direction
yet a low-$T$ transition to the electroweak vacuum still completes (BP~II).

The LISA-observable core is comparably tight: for the $\mathrm{SNR}>10$ points the
$16$--$84\,\%$ chain intervals are $m_H\in[197,222]\,$GeV, $m_A\in[410,431]\,$GeV, $m_{H^\pm}\in[420,437]\,$GeV
in the EWSR subset, widening to $m_H\in[213,322]\,$GeV, $m_A\in[382,451]\,$GeV, $m_{H^\pm}\in[403,457]\,$GeV
once the non-restoring subset is included. The standard-custodial cloud produces the loudest LISA signals of the entire scan. Its
$\mathrm{SNR}>10$ points have $\alpha\simeq0.21\text{--}0.33$ and $\beta/H\simeq50\text{--}145$,
percolating and reheating close together at $T_\text{perc}\simeq T_\text{reh}\simeq44\text{--}50\,$GeV.
The maximal SNR reaches $\simeq141$ ($56$) in \texttt{BSMPT} (\texttt{TL});
restricted to the conservative EWSR subset the \texttt{BSMPT} value drops to $\simeq36$, whereas the
\texttt{TL} maximum is unchanged, as it is attained within that subset.

Crucially, these LISA-observable points are to the largest part
within the HL-LHC reach. The dominant
channel allowing tests of this scenario is resonant $gg\to H\to ZZ$. Figure~\ref{fig:cloudA_zz} shows the predicted effective cross-section
$\sigma(gg\to H)\,\mathrm{BR}(H\to ZZ)$ at $\sqrt s=14\,$TeV against $m_H$, together with the
current and projected $H\to ZZ$ limits and the HL-LHC sensitivity, split into the EWSR (left) and
No EWSR (right) subsets.
Of the LISA-observable realizations, 94.8\,\% (73.8\,\%) of the EWSR (No EWSR) points already lie
above the projected HL-LHC reach in this single channel. The grid of all combinations of
collider channels, used code (\texttt{BSMPT} and \texttt{TL}) and symmetry restoration
requirements can be found in
figure~\ref{fig:cloudAgrid} in appendix~\ref{app:grids}. Combining all collider channels
leaves the EWSR fraction essentially unchanged at $\simeq95\,\%$ ($96\,\%$), based on the GW signal
predictions from \texttt{BSMPT} (\texttt{TL}): $H\to ZZ$ alone
already saturates the reach for the light, SM-like-coupled $H$ of the EWSR subset. Of the other
channels only resonant $A\to Zh\to Z b\bar b$ contributes appreciably, and it does so for the No EWSR
subset: being sensitive to the heavy pseudoscalar rather than to $m_H$, it tests $\simeq45\,\%$
($40\,\%$) of those points and recovers many of the heavier ones that $H\to ZZ$ misses, raising the
combined No EWSR fraction from $\simeq74\,\%$ ($66\,\%$) to $\simeq89\,\%$ ($85\,\%$). The remaining modes,
i.e.~$H\to WW$, $H\to\tau\tau$, di-Higgs $H\to hh$, and the charged-Higgs channels $H^\pm\to\tau\nu$,
$H^\pm \to W^\pm h$, add negligible reach for these points. The di-Higgs channel is the only one
among them which reaches any LISA-observable point at all, testing $3.9\,\%$ ($0.5\,\%$) of the No EWSR
points with $\text{SNR}>10$ in the \texttt{BSMPT} (\texttt{TL}) analysis, all of which sit just above
the $m_H = 251 \, \text{GeV}$ threshold below which no resonant di-Higgs limit exists. Every one of
them is however already covered by $H \to ZZ$ or $A \to Zh$, such that the combined testable
fractions quoted above are unchanged by this channel.

The two subsets differ in a correlated way worth spelling out: The No EWSR subset reaches the
higher SNRs (it contains the global maximum, $\mathrm{SNR}_\texttt{BSMPT}\simeq141$, while the EWSR subset tops out around $36$)
but is the harder one to test in the dominant $H\to ZZ$ channel: its chains extend to heavier $m_H$
(up to $\simeq320\,$GeV), where the resonant $H\to ZZ$ cross section and the search sensitivity both
fall off, so a larger fraction drops below the projected $ZZ$ limit, i.e.~the heavier, lower-lying
population in the right panel of figure~\ref{fig:cloudA_zz}. It is precisely this mass effect that the
complementary $A\to Zh$ channel above partially compensates.

As an independent check and in order to estimate the underlying theoretical uncertainties,
in appendix~\ref{app:uncertainties} we compare,
point by point along the chains, the phase-transition inputs ($T_\text{perc}$, $T_\text{reh}$,
$\alpha$, $\beta/H$, etc.) and the resulting SNR from \texttt{BSMPT} and \texttt{TL}. The two codes agree
on $T_\text{perc}$ at the few-percent level; the leading SNR difference is the sound-shell-lifetime
factor $\mathcal{Y}_\text{sw}$, which shifts the maximum from $\simeq141$ (\texttt{BSMPT}) to
$\simeq56$ (\texttt{TL}). Importantly, the HL-LHC-testable fraction is almost code-independent
($94.8\,\%$ vs.\ $95.9\,\%$), so the complementarity established below does not rest on the GW-modelling
choice. The reason for this is the strong parameter tuning of the GW predictions, required in order to
reach the LISA threshold (and the reason for why an MCMC search is needed in the first place),
whereas the HL-LHC testability is comparably robust across the cloud. Hence, on a point-by-point basis,
the LISA SNR is more sensitive to the underlying theoretical uncertainties than the HL-LHC testability,
but the overall picture of complementarity is robust.

\subsection{GWs and collider-testability in the twisted-custodial cloud}
\label{subsec:cloudB}

The twisted-custodial cloud sits on the other custodial limit, $\lambda_4\simeq-\lambda_5$, where
the heavy CP-even scalar and the charged Higgs are degenerate, $m_H\simeq m_{H^\pm}$. Here $H$ and
$H^\pm$ stay light, $m_H\simeq m_{H^\pm}\simeq170\text{--}200\,$GeV, while the pseudoscalar is pushed
heavy, $m_A\simeq560\text{--}580\,$GeV (see table~\ref{tab:clouds} and appendix~\ref{app:custodial}). As in the standard-custodial cloud the
heavy CP-even scalar stays above the observed Higgs mass, $m_H>m_h$. The MCMC chains comprise in total
$146{,}094$ steps with $\sim144{,}400$ distinct points.

The transitions in this cloud are systematically weaker than in the standard-custodial one:
$\alpha\simeq0.07\text{--}0.12$ (against $0.21\text{--}0.33$) and $\beta/H_*\simeq60\text{--}175$,
with $T_\text{perc}\simeq T_\text{reh}\simeq59\text{--}69\,$GeV. The reason is not that this
cloud releases less vacuum energy, but that it releases it into a hotter plasma: its barrier is on
average about $30\,\%$ lower, so nucleation is reached with less cooling and the
median percolation temperature is about $15\,\%$ higher. Through
$\alpha\propto\Delta\mathcal{F}_0/T_\text{perc}^4$ this alone gives a relative $70\,\%$ decrease in the
transition strengh, see also the $\alpha$--$T_\text{perc}$ panels in
the corner plots in figures~\ref{fig:cornerA_mass} and \ref{fig:cornerB_mass}.
Quantitatively, the bulk of the cloud stays well below detectability, and we use the looser
$\mathrm{SNR}>1$ criterion to characterize its mass range in table~\ref{tab:clouds}. A small No EWSR
subset does reach the LISA threshold in \texttt{TL}, but not in \texttt{BSMPT}: we find seven points with
$\mathrm{SNR}_\texttt{TL}\simeq10\text{--}16$, driven by unusually deep supercooling
($T_\text{perc}/T_\text{crit}\simeq0.3$), whereas
\texttt{BSMPT} keeps the \emph{entire} cloud below the observability threshold of
$\mathrm{SNR}=10$.\footnote{\label{fn:outlier} The
MC chains in fact contained two isolated parameter points which we excluded from the final
analysis, as neither is reproducible under tiny variations of $\lambda_1$ (at the level of relative
changes of $10^{-9}$): one with $\mathrm{SNR}_\texttt{BSMPT} \simeq 25$, the largest
\texttt{BSMPT} value anywhere in this cloud, and one on which \texttt{TL} instead blew up to
$\mathrm{SNR}_\texttt{TL} \simeq 9\,500$ while \texttt{BSMPT} returned $\simeq 3$ for the same
point. With both removed, the largest \texttt{BSMPT} value in the twisted-custodial cloud is
$\mathrm{SNR}_\texttt{BSMPT} \simeq 8.8$. We performed the same tests for the seven points with
$\mathrm{SNR}_\texttt{TL}\simeq10\text{--}16$ and found them to be robust.} These are thus a strongly
parameterisation-dependent, $\mathcal{O}(\text{few})$ effect right at the edge of detectability.
Here, unlike for the testable fractions quoted throughout this work, the differing conventions of
the two codes do matter: the \texttt{TL} values assume four years of observation against
\texttt{BSMPT}'s three, and rescaling them to a common three-year exposure,
$\text{SNR} \propto \sqrt{\mathcal{T}}$, leaves only two of these seven points above the detection
threshold. The foreground included in the \texttt{TL} noise budget acts in the opposite direction, so
this is not the net effect, but it shows that the crossing of the threshold in this cloud sits inside
the difference between the two conventions and should not be read as a robust prediction.
We conclude that the twisted-custodial cloud does not robustly predict
LISA discoveries and only allows for a marginally testable fraction in the most optimistic scenario,
while the standard-custodial cloud is the more optimistic one for a LISA discovery,
with a larger fraction of its points
above the $\mathrm{SNR}=10$ observability threshold.

\begin{figure}
    \centering
    \includegraphics[width=0.85\linewidth]{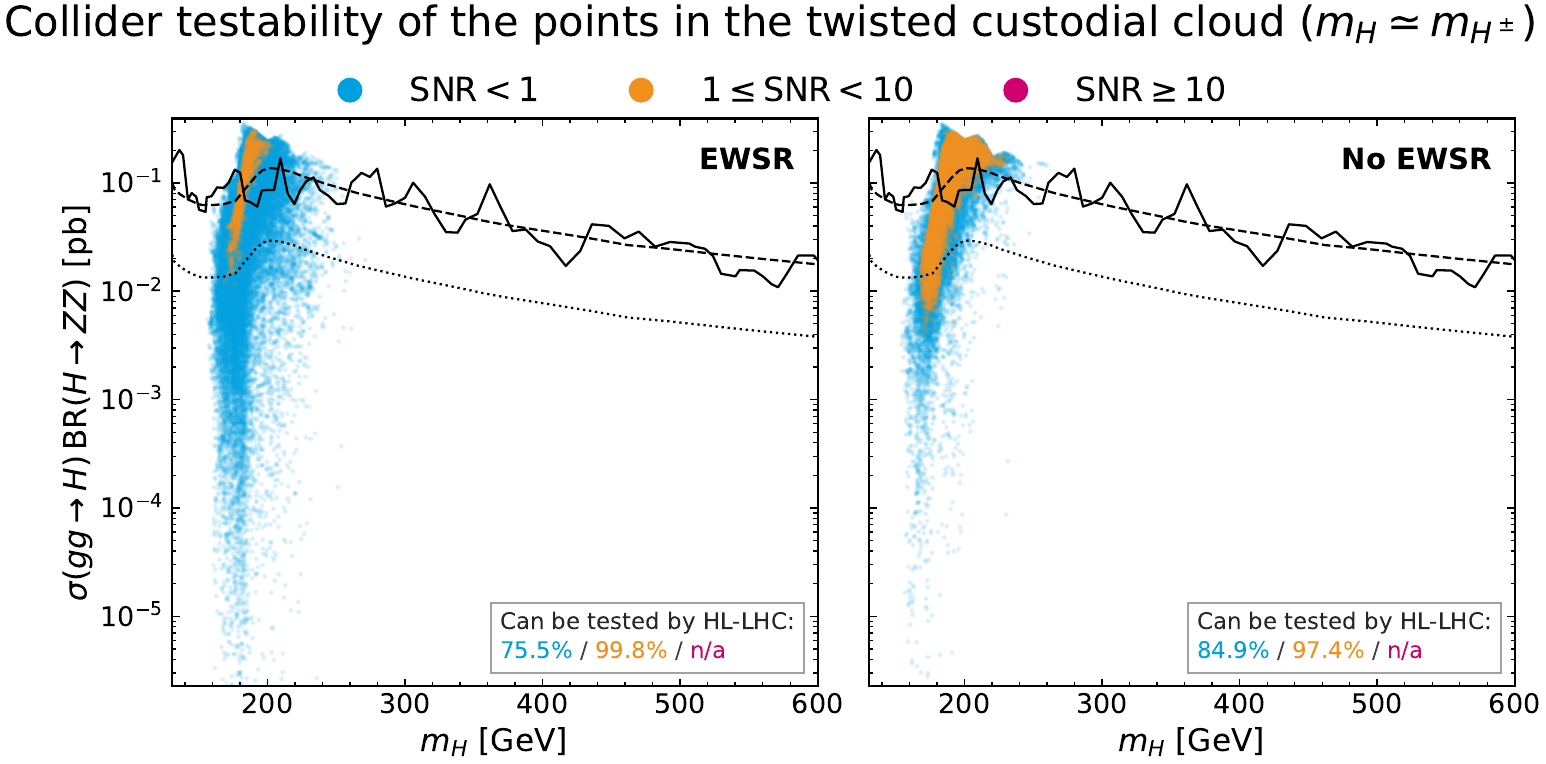}
    \caption{Same as figure~\ref{fig:cloudA_zz}, but for the twisted-custodial cloud.
    Already for FOPTs slightly too weak
    to be observed with LISA, i.e.~$1 \le \text{SNR} < 10$, the collider-testable fraction is
    $99.8\,\%$ (97.4\,\%) when requiring that the electroweak
    symmetry is (not) restored at high temperatures.}
    \label{fig:cloudB_zz}
\end{figure}

The collider picture is the mirror image: the light $H$ makes the twisted-custodial cloud more easily
testable. Figure~\ref{fig:cloudB_zz} shows the same $H\to ZZ$ observable as
figure~\ref{fig:cloudA_zz}; the points just below the LISA detection threshold $1 < \mathrm{SNR} < 10$
cluster tightly at $m_H\simeq170\text{--}200\,$GeV with cross
sections $\sim10^{-2}\text{--}10^{-1}\,$pb, and essentially the entire cloud lies above the projected
HL-LHC sensitivity, with testable fractions near $100\,\%$. Equivalently, the weakest
twisted-custodial transition that the HL-LHC would \emph{miss} corresponds to an SNR below the
$\mathrm{SNR}=1$ floor: the HL-LHC probes every twisted-custodial realization with even a moderate
transition strength, long before LISA or a more sensitive follow-up mission could. The remaining channels and combined fractions are shown in the
grid of figure~\ref{fig:cloudBgrid}, and the coupling-basis correlations in
figures~\ref{fig:cornerB_coupling}
and~\ref{fig:cornerB_mass} (appendix~\ref{app:corner}). We note that
again the $ZZ$ channel which once led to the SM Higgs discovery is the dominant one, also for the twisted-custodial cloud,
with the next-most important channel in our analysis being
$gg \to t b H^\pm$ and subsequent charged Higgs decays $H^\pm \to \tau \nu$, being able to probe between $2.5\,\%$ and $14.5\,\%$,
depending on the electroweak symmetry restoration requirement and the used code (figure~\ref{fig:cloudBgrid}).

The \texttt{BSMPT}-vs.-\texttt{TL} comparison for this cloud
(appendix~\ref{app:uncertainties}) is again tight in $T_\text{perc}$ and
$\alpha$; the SNR ratio scatters slightly more than in the standard-custodial cloud, traceable to the
smaller $\alpha$ and longer sound-shell lifetimes that amplify the differing $\mathcal{Y}_\text{sw}$
prescriptions (section~\ref{subsec:GWB}). The main conclusion regarding the cloud's $\sim100\,\%$ HL-LHC
testability and in general weaker GW signals compared to the standard-custodial cloud
does \textit{not} depend on the code choice, whereas the crossing of the $\mathrm{SNR}=10$ detection
threshold is indeed code-dependent.

\subsection[Comparison with the extremely high SNRs reported in the literature]{Comparison with the extremely high SNRs reported in ref.~\cite{Goncalves:2021egx}}
\label{subsec:goncalves}
As noted in section~\ref{sec:introduction}, ref.~\cite{Goncalves:2021egx} reports SNRs as large as
$10^8$ in regions of the type-I 2HDM parameter space corresponding closely to our standard- and
twisted-custodial clouds, i.e.~six orders of magnitude above the maximal values
$\text{SNR}_\texttt{BSMPT} \simeq 141$ and $\text{SNR}_\texttt{BSMPT}\simeq 8.8$ found in this work
(and $\text{SNR}_\texttt{TL}\simeq 56$ and $\simeq 16$ for the loudest points of the
respective cloud in the \texttt{TL} analysis, which are \emph{not} the same parameter points: the
loudest \texttt{BSMPT} point of the standard-custodial cloud is assigned
$\text{SNR}_\texttt{TL}\simeq 0.6$, and vice versa).
A few further chain points were discarded for the same reason: re-evaluating them at their
tabulated parameters reproduces the \texttt{BSMPT} result rather than the deeply supercooled one
originally stored, so we treat those values as artefacts of the percolation solver. All quoted
maxima are the largest reproducible ones.
Our analysis finds the same qualitative picture as
ref.~\cite{Goncalves:2021egx}, that a strong transition favours light non-standard scalars within
reach of the HL-LHC, but not its specific SNR values. Here we want to argue why we find these
extremely high SNRs to not be realizable in the type-I 2HDM.

Since $\text{SNR}\propto h^2\Omega_\text{GW}\propto K_\text{sw} (\beta/H)^{-2}$
(see eqs.~\eqref{eq:GWspec}--\eqref{eq:SNR}) and the sound wave amplitude is bounded by the kinetic
energy fraction $K_\text{sw}^2<1$,
only a parametrically small $\beta/H$ can raise the SNR by six orders of magnitude.
Ref.~\cite{Goncalves:2021egx} sets the inverse transition duration to the bounce action slope 
$(\beta/H)_{S_3}\equiv T_\text{n}\,\mathrm{d}(S_3/T)/\mathrm{d}T|_{T_\text{n}}$ at
nucleation,\footnote{We further want to point out that in ref.~\cite{Goncalves:2021egx}, nucleation is understood
as the moment in time when $S_3/T \approx 140$, stemming from a heuristic assuming radiation
domination. In the present work, we go beyond this approximation and include the effect of the
equation of state of the electroweak plasma before and after the transition and
compute the GW emission at the time of bubble percolation. We, however, do not expect this
inconsistency to explain the large discrepancy in SNRs between both works.} whereas we take $(\beta/H)_{RH}$
from the mean bubble separation $RH$ at percolation, both in the \texttt{BSMPT} and \texttt{TL} implementation.
As shown in ref.~\cite{Matuszak:2026xsz}, the
bounce action slope $(\beta/H)_{S_3}$ is not a reliable quantity to estimate the inverse
duration for slow transitions,
as it can drop arbitrarily close to zero and can even become negative, $(\beta/H)_{S_3}\gtrsim-\mathcal{O}(100)$,
see also figure~3 in ref.~\cite{Matuszak:2026xsz}. 
The mean bubble separation-based $(\beta/H)_{RH}$, evaluated at percolation, is bounded from below
at $\mathcal{O}(3)$, such that the SNR saturates to more modest values. We therefore suspect the reported
SNRs correspond to $(\beta/H)_{S_3}\to0$, i.e.~arbitrarily strong increases in the GW amplitude,
easily explaining the six orders of magnitude difference.

Along the lines of ref.~\cite{Matuszak:2026xsz} we want to add a physical
reasoning for this discrepancy:
The often-used relation $RH \simeq (8 \pi)^{1/3} / (\beta/H)_{S_3}$
\cite{Caprini:2015zlo,Caprini:2018mtu,Caprini:2024hue,Athron:2023xlk},
used to infer the relevant length scale at which GWs are produced only holds 
at percolation, when corrected by a numerical prefactor of $\mathcal{O}(1)$, and in the limit of very
large $(\beta/H)_{S_3}$. Ref.~\cite{Goncalves:2021egx} remains unclear about their
implementation of the GW signal, but it is reasonable to assume that they in fact used
this relation to obtain GW signal predictions, outside its range of validity. Hence,
a mean bubble separation $RH\gg1$ exceeding the Hubble radius at percolation would
implicitly be assumed in the GW prediction, i.e.~a single bubble filling a larger volume than the
observable universe at percolation, which is obviously inconsistent with the possibility of
bubble collisions within a causally connected region, emitting GWs. As recently shown in
ref.~\cite{Lewicki:2025xyz}, cosmic expansion during ultraslow transitions further is expected
to weaken the $\propto(\beta/H)^{-2}$ scaling, so that the amplitude saturates well
below a naive extrapolation. We hence conclude that the extremely high (and presumably
unbounded) SNRs reported in ref.~\cite{Goncalves:2021egx} cannot be realized in the
type-I 2HDM.

\subsection{Caveats and future directions}
\label{subsec:caveats}

Given the strength of our central claim, in this subsection we want to collect the main caveats
of the performed analysis, grouped by
the stage at which they enter: The MC exploration of parameter space, the GW
signal prediction, and the effective-potential calculation. In each case we argue that
these caveats do not affect the bottom line: strong transitions occupy only small islands
of the 2HDM, essentially all of which the HL-LHC can test.

The first kind of caveats concern the parameter space exploration. Because our chains maximise the
LISA SNR, they settle on the two loudest arms of the $(\Delta m_H,\Delta m_A)$ cross of
ref.~\cite{Goncalves:2021egx}, the standard-custodial cloud ($m_H<m_{H^\pm}\simeq m_A$, left branch)
and the twisted-custodial cloud ($m_H\simeq m_{H^\pm}<m_A$, top branch). Based on our initial
random scan we do not populate the bottom ($m_H\simeq m_{H^\pm}>m_A$), right
($m_H>m_{H^\pm}\simeq m_A$) or central ($m_H\simeq m_{H^\pm}\simeq m_A$) regions where
ref.~\cite{Goncalves:2021egx} still reports $\xi_\text{c}>1$. In fact, we also find marginally
smaller transition strengths in these regions, although below $\xi_\text{c}=1$. As these points carry smaller
mass splittings and hence generally weaker GW signals, cf.~figure~\ref{fig:barrier}, these regions are not
expected to be testable with LISA. This last statement inherits the
limitation of the sampling it rests on. The random scan of section~\ref{subsec:preselection} is the
same one whose failure to reach $\text{SNR}>10$ motivated the MCMC in the first place, and the chains,
seeded inside the two arms, cannot by construction discover a third region. What we establish is
therefore that these are the two loudest regions \emph{reachable by this search strategy}, rather than
a proof that no other corner of the type-I 2HDM hosts a LISA-observable transition. We regard the
former as sufficient for the complementarity claim, since it is the loud regions that the argument is
about, but the distinction should be kept in mind.

On the collider side our reach
rests on (close-to) resonant
$gg\to H\to ZZ$, which ref.~\cite{Goncalves:2021egx} does not consider, whereas we do not
implement the fermionic $H,A,H^\pm\to t\bar t,\,tb$ searches which they found to be
most relevant. The latter is a deliberate choice rather than an omission: the four-top signature
$gg \to t\bar t H/A$ falls short of the corresponding ATLAS limit~\cite{ATLAS:2022rws} by two to
three orders of magnitude in both clouds, the resonant decay $H \to t\bar t$ is kinematically closed
for essentially all model realizations relevant here, and the only unsuppressed channel,
$A \to t\bar t$, is not constrained by the published searches for either cloud. We quantify these
three statements in appendix~\ref{app:grids}, where we also point out that the $A \to t\bar t$
channel may become relevant for the twisted-custodial cloud at full HL-LHC luminosity.
Moreover, our collider reach is projected from published analyses by luminosity scaling,
which is a conservative way to extrapolate: the sensitivity of an LHC search does not in general
improve only as $\sqrt{\mathcal{L}}$, and improved analysis techniques can be expected to strengthen the
constraints beyond what we assume here~\cite{Belvedere:2024wzg}. To the extent that this is so, the
testable fractions we quote are lower bounds.
We assume that the two reach estimates are complementary rather than directly comparable,
allowing for a more comprehensive exploration of the parameter space, possibly also allowing
to extend our statements to model realizations in the standard-custodial branch with $\text{SNR}<10$.
Finally, our scans are restricted to the CP-conserving type-I 2HDM: other Yukawa types (for which the transition
strength is nearly identical~\cite{Goncalves:2021egx}), CP-violating potentials, and
multi-step transitions (e.g.~with CP- or charge-violating intermediate minima~\cite{Aoki:2023lbz})
lie outside the scope of this work and would constitute separate analyses.

The second group of caveats concern the GW prediction. As we argue in appendix~\ref{app:uncertainties},
the two approaches of computing the GW signals based on \texttt{BSMPT} and \texttt{TL} span
the space of possible choices regarding the precise modelling of the signal: The bubble
wall velocity and the lifetime of the soundwave source thereby play a central role,
and we have made sure that the two approaches do not systematically bias the results.
We however want to point out that on a point-by-point basis the SNR can shift by an
$\mathcal{O}(1)$ factor, see also the discussion in appendix~\ref{app:uncertainties}. One additional concern are,
however, outliers like the one point in the twisted-custodial cloud with $\text{SNR}_\texttt{BSMPT}\simeq25$,
discussed in footnote \ref{fn:outlier}: As this point is not reproducible under tiny variations of $\lambda_1$
(at the level of relative changes of $10^{-9}$) or in \texttt{TL}, it is also reasonable to assume that
it (at least partially) led the MC algorithm astray. Given that already all of cloud B with $\text{SNR}>1$ is HL-LHC-testable,
this point is immaterial, however. 

The third kind of caveats concern the effective potential. We resum daisy diagrams in the Arnold-Espinosa
prescription, which dresses only the zero-Matsubara modes that require resummation and is thus
consistent with the power counting of the high-temperature effective theory obtained by dimensional
reduction~\cite{Gorda:2018hvi,Gould:2021oba}; the alternative Parwani scheme instead dresses all modes,
mixing orders in that expansion, and yields systematically larger $\xi_\text{c}$ and more frequent multi-step
transitions~\cite{Basler:2016obg,Lofgren:2023sep,Aoki:2021oez}. The chosen resummation scheme and
renormalization scale
dependence are known theoretical uncertainties, removed only by a full dimensional reduction that we do
not employ~\cite{Croon:2020cgk}. Our Arnold-Espinosa choice is in this sense the better-motivated and
conservative one concerning the SNRs at LISA. Multi-step transitions, prominent mainly under the
Parwani prescription~\cite{Aoki:2021oez}, are accordingly not a relevant feature of the parameter space we find.
Radiatively generated, Coleman--Weinberg-like barriers would require an $\overline{\text{MS}}$-type
renormalization that the on-shell-like scheme adopted here cannot
describe~\cite{Braathen:2025svl,Basler:2018cwe};
unlike the strongly-transitioning single-field conformal $U(1)$ case~\cite{Bringmann:2026xcx},
multi-field conformal potentials such as the 2HDM do not generically produce strong
transitions~\cite{Benincasa:2026dhg}, so this regime is not expected to host
additional LISA-observable points. On a more fundamental level, however, the concern brought forward
in ref.~\cite{Kainulainen:2019kyp} about the breakdown of the perturbative expansion in the 2HDM
for large couplings, as required for strong transitions, remains. Consequently, a non-perturbative treatment
would be warranted to fully confirm the results of this work. It remains to be seen how large-scale
parameter scans and lattice simulations can be combined in a single framework to answer this question.
Given previous results on the difference of perturbative and non-perturbative treatments of the
electroweak phase transition in the SM and its extensions~\cite{Kajantie:1995kf,Laine:2000rm},
we expect that the qualitative picture of two narrow clouds of strong transitions will remain
unchanged, whereas the precise SNR predictions of individual model realizations will most
likely shift~\cite{Lewicki:2024xan}.
Similar to the discussion in appendix~\ref{app:uncertainties},
we expect that the HL-LHC-testable fraction of the LISA-observable points will remain robust under
such a shift in the LISA-observable part of the parameter space, as the HL-LHC testability is far
less sensitive to the precise parameter tuning than the one required to reach the LISA threshold.

\section{Conclusion}
\label{sec:conclusion}

In this work we have performed a combined analysis of the GW signals from strong
first-order electroweak phase transitions in the type-I 2HDM and their complementarity with collider
searches at the HL-LHC. Using a combination of a random scan and a targeted MCMC search,
cross-validated between \texttt{BSMPT v3} and \texttt{TransitionListener v2}, we find that strong
phase transitions with sizeable LISA SNRs are realized only in two narrow, well-separated regions of
parameter space, the standard-custodial cloud and the twisted-custodial cloud, characterized by near-degeneracies $m_A \simeq m_{H^\pm}$ and
$m_H \simeq m_{H^\pm}$, respectively (table~\ref{tab:clouds}). This confirms that strong first-order
electroweak phase transitions remain difficult to realize in the type-I 2HDM, and
are confined to comparatively small islands of the available parameter space.

Of these two regions, only the standard-custodial cloud produces signals exceeding the LISA detectability threshold of
$\text{SNR} > 10$, with maximal values of $\text{SNR} \simeq 141$ for $m_H \simeq
180\text{--}250\,\text{GeV}$ and $m_A \simeq m_{H^\pm} \simeq 400\text{--}450\,\text{GeV}$. The twisted-custodial cloud, with $m_H
\simeq m_{H^\pm} \simeq 170\text{--}200\,\text{GeV}$ and $m_A \simeq 560\text{--}580\,\text{GeV}$, remains
effectively below this threshold (see section~\ref{subsec:cloudB}). In
section~\ref{subsec:goncalves} we argued that the much larger SNRs of up to $10^8$ reported for similar
parameter-space regions in ref.~\cite{Goncalves:2021egx} are not physically attainable in the 2HDM:
they would require $\beta/H \to 0$, which is in tension both with the resulting mean bubble
separation exceeding the Hubble radius and with the validity of current simulation-based GW spectral shapes.

On the collider side, we find that the SM-like Higgs discovery channel $H \to ZZ$ is again the most
powerful probe of both clouds at the HL-LHC, and we want to motivate further studies of it for
$m_H \lesssim 200\,\text{GeV}$, i.e. around the $H\to ZZ$ resonance. The HL-LHC
is expected to probe the
vast majority of the LISA-observable parameter space in the standard-custodial
cloud ($95\,\%$--$96\,\%$, depending
on the code used for the phase-transition computation) as well as essentially
all of the twisted-custodial cloud, well
before LISA collects data. We find that dropping the condition of electroweak symmetry restoration 
at high temperatures opens up the available model parameter space, leading to
stronger possible LISA signals at the cost of a decreased reach of the collider searches
($85\,\%$--$89\,\%$ in the standard-custodial cloud and again virtually the whole of the
twisted-custodial cloud).

Our findings establish a striking complementarity between the two experiments:
if the HL-LHC excludes an $H \to ZZ$ signal in the relevant mass window, this will rule out the
corresponding strong-phase-transition scenarios as a source of an observable LISA signal.
Conversely, the discovery of such a signal, in particular for $\tan\beta \simeq 14$ characteristic
of the standard-custodial cloud,
would constitute strong motivation for a dedicated LISA analysis of the corresponding GW
background, including effects such as the bubble wall velocity that were treated only at the level
of an estimated theoretical uncertainty in this work (appendix~\ref{app:uncertainties}).

Taken together, our results illustrate that GW and collider experiments offer highly
complementary probes of extended scalar sectors: while LISA is sensitive to the dynamics of the
electroweak phase transition itself, the HL-LHC can independently test the same regions of parameter
space through the spectrum and couplings of the additional scalars. The interplay between these two
avenues, exemplified here for the type-I 2HDM, motivates similar combined analyses for other
extensions of the scalar sector.

\acknowledgments

We thank Lisa Biermann for advice on the use of \texttt{BSMPT v3}, and Jonas
Matuszak for support in the development of a 2HDM model file for \texttt{TransitionListener v2}.
We are further grateful to Atri Dey, Rikard Enberg, Georg Weiglein and Johannes Braathen
for discussions on the 2HDM and respective experimental limits.
We thank Harri Waltari for actual collaboration in the initial stages of this research as 
well as for innumerable discussions throughout.
SM is supported in part through the NExT Institute and STFC Consolidated Grant ST/X000583/1.
CT acknowledges support from a Feodor Lynen Fellowship of the Alexander von Humboldt Foundation
and is supported by
the Spanish National Grant PID2022-137268NA-C55 and Generalitat Valenciana through the
grant CIPROM/22/69. Computations for this work were performed on the local SOM
clusters Gluon and Graviton at the Instituto de Física Corpuscular (IFIC),
as well as on computational resources provided by the National Academic Infrastructure
for Supercomputing in Sweden (NAISS), funded by the Swedish Research Council.

\appendix

\section{Custodial symmetry in the 2HDM}
\label{app:custodial}

This appendix works out the custodial symmetry that underpins the electroweak-precision
discussion of section~\ref{sec:paramspace} and the two clouds of
section~\ref{sec:results}: why the tree-level relation $\rho=1$ is protected by such a symmetry,
why the 2HDM admits two inequivalent such symmetries, and why these pick out the two coupling
relations $\lambda_4\simeq\pm\lambda_5$.

A complex doublet $\Phi=(\varphi^+,\varphi^0)^\text{T}$ carries four real fields, on which the
Higgs kinetic term and potential possess a global $SU(2)_L\times SU(2)_R$ symmetry larger than the
gauged $SU(2)_L\times U(1)_Y$. It is made manifest by arranging $\Phi$ and its conjugate
$\tilde\Phi=\mathrm{i}\sigma_2\Phi^*$ into the $2\times2$ bi-doublet
\begin{align}
    M = (\tilde\Phi,\,\Phi) = \begin{pmatrix}\varphi^{0*}&\varphi^+\\-\varphi^-&\varphi^0\end{pmatrix},
    \qquad \varphi^-\equiv(\varphi^+)^*,
\end{align}
which transforms as $M\to L\,M\,R^\dagger$, with $L\in SU(2)_L$ and $R\in SU(2)_R$. Here
$L$ is the ordinary, gauged weak isospin acting from the left, while $R$ is a \emph{global}
$SU(2)_R$ acting from the right, of which the Standard Model gauges only the third generator, 
i.e.~hypercharge $U(1)_Y\subset SU(2)_R$. The electroweak vacuum $\langle M\rangle=(v/\sqrt2)\,\mathbf{1}$
is proportional to the identity and is therefore invariant only under the diagonal subgroup $L=R$.
This unbroken $SU(2)_V$ is the \emph{custodial} symmetry: it rotates the three gauge bosons
$W^{1,2,3}$ as a degenerate triplet, so that at tree level $m_W=m_Z\cos\theta_W$, i.e.\ $\rho=1$. The
symmetry is only approximate even in the Standard Model: hypercharge ($g'\neq0$) and the $t$--$b$ mass
splitting break $SU(2)_R$ explicitly and shift $\rho$ from unity at one loop. It is convenient to
measure any further, new-physics deviation relative to this Standard-Model value, and this is what the
oblique parameter $T$ (equivalently $\Delta\rho$ below) quantifies.

With two doublets there are two bi-doublets $M_1,M_2$; the gauge and kinetic terms still respect
the global $SU(2)_R$, and whether the scalar potential does decides whether the extra scalars shift
$\rho$ from unity. The second doublet need not transform in the same way as the first, and there are
two inequivalent consistent choices,
\begin{align}
    \text{standard:}\quad M_2\to L\, M_2\, R^\dagger\,, \qquad\qquad
    \text{twisted:}\quad M_2\to L\, M_2\,\sigma_3\, R^\dagger\,\sigma_3\,,
\end{align}
the second being the statement that $M_2\sigma_3$, rather than $M_2$, transforms in the standard
way. The extra $\sigma_3=\mathrm{diag}(1,-1)$ acts in the custodial $SU(2)_R$ space and commutes with
the gauged hypercharge generator, so both choices are compatible with the Standard-Model gauging; its
effect is to exchange the CP-even and CP-odd neutral scalar that partners the charged Higgs in a
custodial multiplet. This is the twisted custodial symmetry of
refs.~\cite{Gerard:2007kn}.

The bilinears appearing in the potential are exactly the invariants of these two symmetries. A
short computation gives
\begin{align}
    \mathrm{Tr}(M_1^\dagger M_2) = 2\,\mathrm{Re}\,\Phi_1^\dagger\Phi_2\,, \qquad
    \mathrm{Tr}(M_1^\dagger M_2\,\sigma_3) = -2\mathrm{i}\,\mathrm{Im}\,\Phi_1^\dagger\Phi_2\,.
\end{align}
The first trace is invariant when both doublets rotate identically ($R$ cancels between
$M_1^\dagger$ and $M_2$), so $\mathrm{Re}\,\Phi_1^\dagger\Phi_2$ is the \emph{standard}-custodial
invariant; the second is invariant only when $M_2$ carries the $\sigma_3$ twist, so
$\mathrm{Im}\,\Phi_1^\dagger\Phi_2$ is the \emph{twisted}-custodial invariant. In these variables the
$\lambda_4$, $\lambda_5$ part of the scalar potential is exactly the sum of their squares,
\begin{align}
    \label{eq:lambda45}
    V_\text{tree} \supset \lambda_4\,|\Phi_1^\dagger\Phi_2|^2 + \frac{\lambda_5}{2}\big[(\Phi_1^\dagger\Phi_2)^2+\mathrm{h.c.}\big]
    = (\lambda_4+\lambda_5)\big[\mathrm{Re}\,\Phi_1^\dagger\Phi_2\big]^2
    + (\lambda_4-\lambda_5)\big[\mathrm{Im}\,\Phi_1^\dagger\Phi_2\big]^2 \, .
\end{align}
The potential therefore respects a custodial symmetry (and leaves $\rho=1$ unperturbed at tree
level) only when the
coefficient of the \emph{other} invariant vanishes, i.e.\ for
$\lambda_4=\pm\lambda_5$. For $\lambda_4=\lambda_5$ (the \emph{standard} limit) the twisted
invariant drops and the
charged Higgs is degenerate with the pseudoscalar~\cite[eq.~(11)]{Gunion:2002zf},
\begin{align}
    \label{eq:custodial}
    m_A^2-m_{H^\pm}^2 = \tfrac{1}{2}(\lambda_4-\lambda_5)\,v_\text{EW}^2
    \;\xrightarrow{\;\lambda_4=\lambda_5\;}\; m_{H^\pm}=m_A \, ,
\end{align}
whereas for $\lambda_4=-\lambda_5$ (the \emph{twisted} limit) the standard invariant drops and
$H^\pm$ pairs with the heavy CP-even scalar instead. The two limits are not quite symmetric,
however. Equation~\eqref{eq:custodial} involves the quartics alone, so $\lambda_4=\lambda_5$
enforces $m_{H^\pm}=m_A$ by itself. The corresponding CP-even splitting also picks up the
mixing of the two doublets~\cite[eqs.~(10)--(13)]{Gunion:2002zf},
\begin{align}
    \label{eq:custodial_twisted}
    m_H^2-m_{H^\pm}^2 = \Big[\tfrac{1}{2}(\lambda_4+\lambda_5)
      + s_\beta^2 c_\beta^2\,(\lambda_1+\lambda_2-2\lambda_{345})\Big]v_\text{EW}^2 \, ,
\end{align}
in the alignment limit, so that $\lambda_4=-\lambda_5$ removes only the first term. The twisted
degeneracy $m_{H^\pm}=m_H$ therefore requires either $\lambda_1+\lambda_2=2\lambda_{345}$ in
addition, or a large $\tan\beta$, for which $s_\beta^2c_\beta^2\to0$. Neither holds exactly in the
sampled twisted-custodial cloud: there $\lambda_4+\lambda_5\simeq-0.6$ and
$s_\beta^2c_\beta^2(\lambda_1+\lambda_2-2\lambda_{345})\simeq0.37$, two contributions of opposite
sign and comparable size whose partial cancellation leaves a residual
$m_H-m_{H^\pm}\simeq12\,$GeV. What holds the cloud together is thus not an exact symmetry of the
potential but the electroweak fit discussed next, which tolerates precisely such a residual.

Away from these limits the extra scalars break the custodial
symmetry and give a one-loop contribution to $\rho$ \emph{on top of} the Standard-Model value; in the
alignment limit relevant here this new-physics piece reads~\cite{Grimus:2007if,Dorsch:2013wja}
\begin{align}
    \label{eq:deltarho}
    \Delta\rho = \frac{1}{32\pi^2 v_\text{EW}^2}\big(\,F_{H^\pm,A}+F_{H^\pm,H}-F_{A,H}\,\big)\,, \qquad
    F_{x,y}=\frac{m_x^2+m_y^2}{2}-\frac{m_x^2 m_y^2}{m_x^2-m_y^2}\log\frac{m_x^2}{m_y^2}\,,
\end{align}
with $F_{x,x}=0$. Each $F_{x,y}$ vanishes when the two states are degenerate, so $\Delta\rho$
vanishes in exactly the two custodial limits above, $m_{H^\pm}=m_A$ or $m_{H^\pm}=m_H$. This is the
quantity constrained by the oblique parameter $T$, to which it is directly proportional,
$\Delta\rho=\alpha_\text{em}\,T$. The electroweak fit allows only $|T|\lesssim0.1$~\cite{ParticleDataGroup:2022pth},
i.e.\ $|\Delta\rho|\lesssim10^{-3}$; for scalar masses of a few hundred GeV this caps the
custodial-violating splitting at $\mathcal{O}(10)\,$GeV, the width of the two arms in
figure~\ref{fig:preselection}. These are the standard- and twisted-custodial clouds explored in
section~\ref{sec:results}.

\section{Overview of all studied collider signatures}
\label{app:grids}

\begin{figure}
    \centering
    \includegraphics[width=\linewidth,height=0.80\textheight,keepaspectratio]{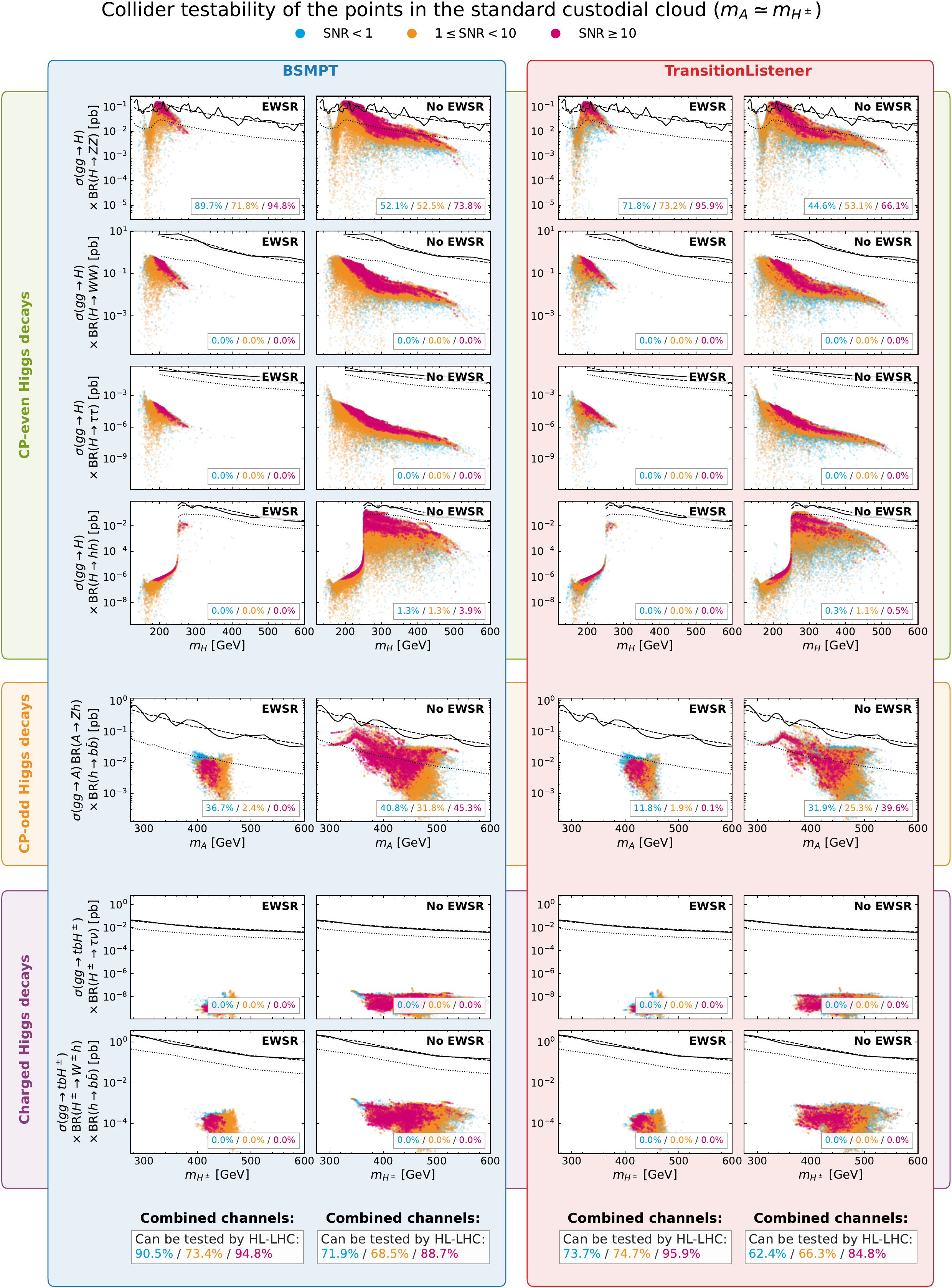}
    \caption{Collider testability of the points of the standard-custodial cloud across all search
    channels considered. Each row is one channel, the left and right blocks show the
    \texttt{BSMPT} and \texttt{TL} SNR,
    each split into the EWSR and No EWSR subsets.
    Points are coloured by SNR bin; solid and dashed curves are current and projected
    exclusion limits and the dotted curve the projected HL-LHC sensitivity. The boxed percentages
    give the HL-LHC-testable fraction in each SNR bin, per channel and combined.}
    \label{fig:cloudAgrid}
\end{figure}

\begin{figure}
    \centering
    \includegraphics[width=1\linewidth]{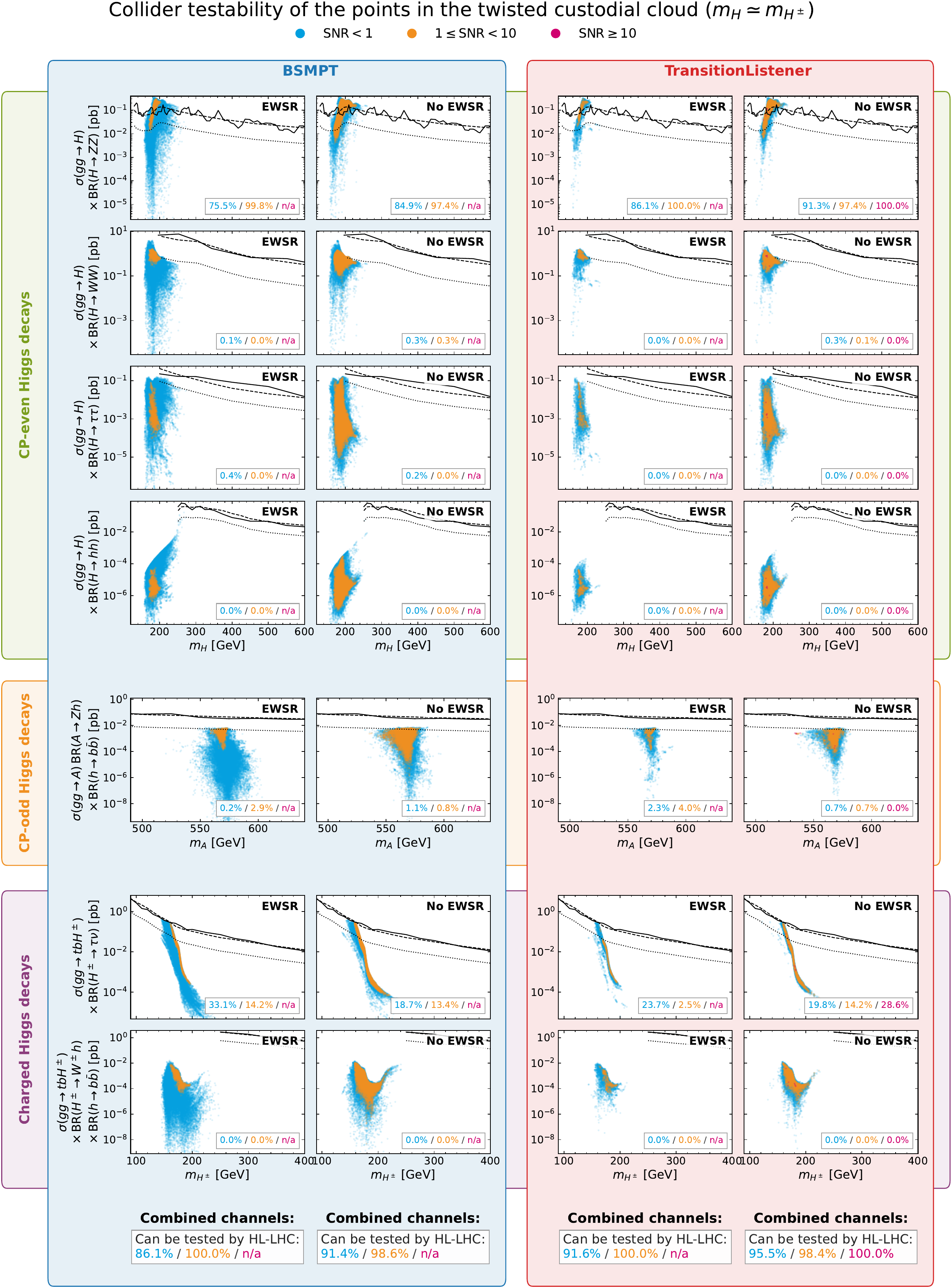}
    \caption{Same as figure~\ref{fig:cloudAgrid}, for the twisted-custodial cloud.}
    \label{fig:cloudBgrid}
\end{figure}

In figures~\ref{fig:cloudAgrid} and \ref{fig:cloudBgrid} an overview of the collider
testability of the BSM scalar decays is presented. Figure~\ref{fig:cloudAgrid} presents
the results for the standard custodial cloud ($m_A \simeq m_{H^\pm}$), whereas
figure~\ref{fig:cloudBgrid} regards the twisted custodial cloud ($m_H \simeq m_{H^\pm}$).
Each of the two plots is organized in a grid shape: The different decay channels, grouped
in decays of the CP-even Higgs $H$ (green), the CP-odd Higgs $A$ (orange), and the charged
Higgs $H^\pm$ (purple), are the rows; the columns correspond to the sampled parameter
points within the given cloud, as computed by \texttt{BSMPT} (blue) or \texttt{TL} (red).
Further, in each of the said two columns, a distinction between points which do and do not
restore the electroweak symmetry at high $T$ is made. In each of the panels, the effective
cross section for the corresponding row's process is printed on the vertical axis, while
the respective decaying scalar mass is on the horizontal axis. The solid (dashed) lines
correspond to the observed (expected) upper limits using LHC data as described in
section~\ref{sec:paramspace}. The used HL-LHC expected upper limits, resulting from a
rescaling of their LHC counterparts, are shown as dotted lines. Further, each panel
includes colour-coded information on the fraction of points with an
$\text{SNR} \ge 10$ (pink), $1 \le \text{SNR} < 10$ (orange)  and $\text{SNR} \le 1$ (cyan)
which can be tested at the HL-LHC. A bottom row combines the collider testability in these
SNR ranges for the selection of sampled points in the respective row. Points which are
beyond the mass range of the shown limits are considered not testable.

We now want to discuss the main information visible in figure~\ref{fig:cloudAgrid}: The most
relevant channel for testing the high-SNR region of the standard custodial cloud
(regardless of which of the two codes is used and whether one imposes a cut concerning
the EWSR) is the $H \to ZZ$ channel (see figure~\ref{fig:cloudA_zz} in the main text for a
zoomed-in version of these panels). For the EWSR cases, no other channel contributes
meaningfully to the collider testability of these model realizations and \texttt{BSMPT}
and \texttt{TL} agree on roughly 95--96\,\% of the sampled points with
$\text{SNR} \ge 10$ being testable at the HL-LHC.\footnote{We point out that the difference
between the testability fractions for $\text{SNR} \le 1$ between \texttt{BSMPT} and
\texttt{TL} are due to the two codes being evaluated on
different sets of points. We did not rerun all points through \texttt{TL},
but only those which already passed an $\text{SNR}_\texttt{BSMPT} > 0.1$ threshold, and of
those \texttt{TL} returns a GW prediction only for a part, mostly because
it finds that the transition does not complete. We quantify this in
appendix~\ref{app:uncertainties}.} All other decay channels remain far below the current
reach of the LHC and can only marginally improve the overall reach of our central claim.
For the No EWSR columns, the distribution of points is qualitatively similar. A notable
difference is that the $A \to Z h$ channel becomes important for constraining this scenario,
whereas the other channels still contribute virtually nothing to the total reach of the
HL-LHC testability fractions. Based on these results, the most relevant collider constraints
will be the $H \to ZZ$ channel for $2 m_Z < m_H \lesssim 300 \, \text{GeV}$ and the
$A \to Zh$ channel for $300 \, \text{GeV} < m_A \lesssim 600 \, \text{GeV}$. The biggest
potential influence on the reach of our claims would be an $\mathcal{O}(1)$ improvement
of the $H \to ZZ$ searches, which would suffice to close the parameter space for
standard-custodial and LISA-visible phase transitions in the type-I 2HDM. 

The collider testability grid presented in figure~\ref{fig:cloudBgrid} for the twisted
custodial cloud has similar features as the standard custodial one presented above:
Again, the most constraining channel is the $H \to ZZ$ one, now however with the biggest
difference being the SNR values which can be reached at LISA within that family of model
realizations: The SNRs saturate at values below 10, which we consider as the detection
threshold in this work. In the \texttt{BSMPT} dataset none of the points has an SNR above
threshold, whereas a small number of points in the No EWSR \texttt{TL} set allows for a
detection with LISA. Still, all of the points with $1 \le \text{SNR} < 10$ (regardless of
the transition code used) can be tested at the HL-LHC. Dropping the EWSR restriction hereby
only opens the parameter space slightly, and the scalar masses favoured by loud GW signals
are still focused in small ranges. Notably, the CP-even Higgs mass $m_H$ lies in a very
short range just around twice the electroweak gauge boson masses, making the $H \to WW$
channel a promising future candidate, if it could be extended to smaller masses than
$200 \, \text{GeV}$. Interestingly, the associated charged Higgs production through
$gg \to tbH^\pm$ with subsequent decays $H^\pm \to \tau \nu$ allows to close the parameter
space of those model realizations with $1 \le \text{SNR} < 10$, which cannot be tested
through $H \to ZZ$. The $A \to Zh$ channel which was found relevant for constraining the
LISA-visible points in the standard-custodial cloud here shows to be only of subdominant
importance due to the increased $A$ mass.

Neither grid contains a row for a top-quark pair in the final state. Ref.~\cite{Goncalves:2021egx}
identifies these channels as the most promising HL-LHC probes of strong transitions, and we
therefore quantify them here: none of the three constrains the LISA-observable region of either
cloud, each for a different reason. The four-top signature
$gg \to t\bar{t}H/A \to t\bar{t}t\bar{t}$ is too small by two to three orders of magnitude
compared to the expected limits: the
predicted rates peak at $\sigma \times \text{BR} \simeq 4 \times 10^{-5} \, \text{pb}$ in the
standard-custodial and $10^{-4} \, \text{pb}$ in the twisted-custodial cloud, against an observed
limit of $1.5 \times 10^{-2} \, \text{pb}$ at $m_{H/A} = 400 \, \text{GeV}$~\cite{ATLAS:2022rws}, a
gap which the luminosity rescaling of section~\ref{subsec:hllhc} does not close. The resonant decay
$H \to t\bar{t}$ is kinematically closed throughout: no twisted-custodial point with
$\text{SNR} > 1$ and only $10\,\%$ of the standard-custodial points with $\text{SNR} > 10$ have
$m_H > 2 m_t$, the latter forming the heavy tail of the No EWSR subset where the rate does not
exceed $10^{-3} \, \text{pb}$. Only $A \to t\bar{t}$ is unsuppressed, and for it no
model-independent limit on $\sigma \times \text{BR}$ exists, since the signal interferes with the
$t\bar{t}$ continuum and appears as a peak-dip rather than as a
resonance~\cite{ATLAS:2024vxm}. The corresponding search
is instead interpreted in the $(m_A, \tan\beta)$ plane for a type-II 2HDM in the alignment limit
with mass-degenerate $m_A = m_H$~\cite{ATLAS:2024vxm}. Neither cloud is excluded by that
contour. The search constrains small $\tan\beta$, where the
production rate $\sigma(gg \to A) \propto \cot^2\beta$ is largest, so the relevant measure is how
far above the exclusion boundary in $\tan\beta$ our points sit. For the LISA-observable points of
the standard-custodial cloud that distance is a factor of $2.4$, which by the $\cot^2\beta$ scaling
corresponds to a production rate lower than the excluded one by $2.4^2\simeq5.8$. Two features of our parameter space weaken the
constraint further: $H$ lies below the $t\bar{t}$
threshold in both clouds, so that only $A$ contributes instead of two degenerate states, and the
open $A \to Zh$ and $A \to ZH$ decays dilute $\text{BR}(A \to t\bar{t})$ to below $2\,\%$.
A cleaner handle is provided by the recent CMS search of ref.~\cite{CMS:2025dzq}, which quotes
upper limits on the coupling of the resonance to the top quark rather than on a cross section, and
which therefore applies to the type-I model without the detour through a type-II benchmark: this
coupling is $\cot\beta$, of which our clouds realize $0.07$ and $0.31$ in the median. Their limits
are given for relative widths between $0.5\,\%$ and $25\,\%$, a range which the pseudoscalars of the
standard-custodial cloud do populate, with a median total width of $1.4\,\%$ of $m_A$. The twisted-custodial
cloud, however, clears the same boundary by only $19\text{--}36\,\%$ in $\tan\beta$, such that this
channel may well develop into a second, independent HL-LHC probe of that cloud once the full
luminosity is collected. Since the twisted-custodial cloud is already testable in its entirety
through $H \to ZZ$, this would corroborate rather than modify our conclusions.

\section{Theoretical uncertainties of the GW signal prediction}
\label{app:uncertainties}

In this appendix we discuss and quantify the theoretical uncertainties on the GW signal prediction that arise
from the modelling choices entering the analysis of section~\ref{sec:GW-2hdm}--\ref{sec:results}. We first
list the expected sources of uncertainty and then discuss the two dominant ones in turn: the residual
differences between \texttt{BSMPT~v3} and \texttt{TransitionListener~v2} in the macroscopic transition
parameters, and the bubble-wall-velocity treatment. Neither of them alters the main result of this
work, i.e.~the HL-LHC--LISA complementarity in the standard-custodial cloud and the twisted-custodial
cloud, which we briefly recap at the end of this appendix.

Each step in the prediction of the GW signal starting from a microscopic description of the 2HDM in
terms of the Lagrangian parameters in eq.~\eqref{eq:coupling_basis} and further the evaluation of a
given stochastic GW background's observability with LISA introduces uncertainties. As of now, a full
quantitative computation of the resulting uncertainty bands for a predicted $h^2\Omega_\text{gw}(f)$
and the corresponding discrepancy in the $\text{SNR}_\text{LISA}$ is still outstanding, see however
refs.~\cite{Croon:2020cgk, Guo:2021qcq} for previous work in these regards. Here we want to provide
the reader with a list of approximations and assumptions used throughout this work. Whenever possible,
we try to quantify the resulting uncertainty and their impact on our central claim.

The first source of uncertainty appears in the modelling of the bubble nucleation rate, which itself
depends on a specific choice for the effective potential. In this work we fixed the renormalization 
scale $\bar{\mu}$ to the electroweak scale $v_\text{EW} = 246 \, \text{GeV}$ and specifically chose
the Landau gauge to allow for an easy separation of unphysical Goldstone and physical Higgs degrees
of freedom. Moreover, we chose to work within the imaginary time formalism in 4D with Debye masses
being resummed through the Arnold-Espinosa prescription. In ref.~\cite{Croon:2020cgk} it has been
shown that each of these assumptions can yield orders of magnitude uncertainties for
$h^2\Omega_\text{gw}(f)$ for a given point in model parameter space. Recently it has been shown, 
however, that these uncertainties only appear amplified due to the strong dependence of the GW
amplitude on the transition temperature, such that the solution of the inverse problem (i.e.
finding the parts of  model parameter space explaining a given signal) only features corresponding
uncertainties at the level of $\mathcal{O}(10\,\%)$. This is in line with previous
studies~\cite{Lewicki:2024xan}. The up-to-$\mathcal{O}(10^3)$ uncertainty in the GW prediction hence
reduces to a mild $\mathcal{O}(10^{-1})$ uncertainty in the model parameters due to the strong
degree of tuning required to generate observable GW signals within the 2HDM. The latter also became
apparent in our analysis in form of the need of elaborate parameter exploration techniques to search
for the parts of parameter space which feature $\text{SNR}_\text{LISA} \ge 10$. Since the collider
phenomenology only depends on the mass spectrum and not on specific parametric cancellations in the
2HDM mass spectrum, we argue that a similar conclusion can be made for the case at hand: A more
rigorous computation of the GW signal like the one presented in ref.~\cite{Kierkla:2026bnm} would
find large discrepancies on a point-to-point basis, but find the same LISA-testable features in the
2HDM parameter space at only slightly shifted parameter values, such that the collider constraints
and hence our central claim would remain virtually unaffected.

In a similar fashion, and to some extent showing the validity of the previous argument, also the
following sources of theoretical uncertainty can be argued to be relevant in a comparison of given
(tuned) parameter choice's GW prediction, but irrelevant for the central claim we want to make: In
our comparison of \texttt{BSMPT} and \texttt{TL}, two separate sets of assumptions
were used to compute the GW signal. While \texttt{BSMPT} assumes a radiation-like equation-of-state
in the fluids in both phases ($c_\text{s} = 1 / \sqrt{3}$), \texttt{TL} computes the
speed of sound in both phases. Moreover, \texttt{TL} unlike \texttt{BSMPT} uses the
equation-of-state parameter in the symmetric phase to evaluate the time-temperature relation present
in the integrals over the bubble nucleation rate needed to compute the true-vacuum fraction's time
dependence or the mean bubble separation. Furthermore, the used bubble wall velocity in both codes
disagrees (with \texttt{BSMPT} computing the wall velocity to be $v_\text{w} = 0.8\text{--}0.9$ and
\texttt{TL} almost always setting it to $v_\text{w} = 1$ after a cross-check with
the LTE approximation). Moreover, in \texttt{BSMPT} the reheating temperature is computed in the
instantaneous reheating approximation and assuming a constant number of degrees of freedom, while
\texttt{TL} drops these assumptions. Concerning the modelling of the lifetime of
the soundwave sources, described by the factor $\mathcal{Y}_\text{sw}$, \texttt{BSMPT} uses an
analytical solution based on the assumption of radiation domination in the broken phase, whereas
\texttt{TL} uses a model-independent (and arguably less accurate) expression for
the same quantity. The overall result is that \texttt{TL} predicts $\alpha$
($(\beta/H)_{RH}$) to be $10\text{--}15\,\%$ smaller (larger) than predicted by \texttt{BSMPT}, even
though the percolation and reheating temperatures match at the $\mathcal{O}(1 \, \%)$-level, similar
to the effects expected mentioned in ref.~\cite{Croon:2020cgk}.
The apparent disagreement in the inverse duration is, however, almost entirely a matter of
convention rather than of code. \texttt{BSMPT} reports the bounce-action slope $(\beta/H)_{S_3}$
while using $(\beta/H)_{RH}$ internally, and comparing like with like the two codes agree closely:
over the points for which both return a value, the ratio of the two $S_3$-based estimates has a
median of $1.03$ in either cloud, with a Spearman rank correlation of $0.96$ (standard-custodial) and $0.997$
(twisted-custodial). The quoted $10\text{--}15\,\%$ is instead the offset between the two definitions inside
\texttt{TL} itself, $(\beta/H)_{RH}$ exceeding $(\beta/H)_{S_3}$ by a median $14\,\%$
and $9\,\%$ in the two clouds. It should therefore not be read as a cross-code uncertainty on the
inverse duration; the genuine cross-code spread on that quantity is at the few-per-cent level. Due to the aforementioned changes
in the macroscopic phase transition description, we find a median suppression $\text{SNR}_\text{TL}
\simeq 0.3 \,  \text{SNR}_\text{BSMPT}$ of the predicted SNR and long tails of the distribution
for the ratio of both quantities. Nevertheless, the analysis presented in section~\ref{sec:results}
concludes that the central claim is independent of the code used in our analysis. This is due to the
same argument as presented in favour of the 4D effective potential in the Arnold-Espinosa scheme:
Even though the precise position of the SNR maximum in the 2HDM parameter space depends on the used
code, the shift is so minuscule that it does not affect our overall conclusion.

The comparison can only be made where both codes converge, and
\texttt{TL} does not return a GW prediction for every realization for which
\texttt{BSMPT} does. Of the points with $\text{SNR}_\texttt{BSMPT} > 0.1$, which are the ones passed
to it, it yields an SNR for $71.7\,\%$ in the standard-custodial and $43.9\,\%$ in the
twisted-custodial cloud, and for $62.4\,\%$ of the LISA-observable band
$\text{SNR}_\texttt{BSMPT} > 10$. The shortfall is mostly physical rather than numerical: in
$97\,\%$ ($87\,\%$) of the standard-custodial (twisted-custodial) failures the code reports that the
transition does not complete --- no nucleation within a Hubble time, no percolation temperature, or a
false-vacuum fraction that never drops below $1\,\%$ --- and only the remaining $3\,\%$ ($13\,\%$)
are genuine numerical failures. This traces back to the more careful treatment of the percolation
criterion and of the equation of state~\cite{Matuszak:2026xsz}: part of what \texttt{BSMPT} counts as
a completed first-order transition does not complete once these effects are included. The two codes
are therefore compared on the intersection of the points on which both converge, and the
\texttt{TL} panels of figures~\ref{fig:cloudAgrid} and~\ref{fig:cloudBgrid} contain
correspondingly fewer points.
These statements rest on a point-by-point comparison of both codes along the chains of the two
clouds, in which the leading residual scatter of the SNR ratio is induced by the soundwave-lifetime
factor $\mathcal{Y}_\text{sw}$ of eq.~\eqref{eq:GWspec}. It lowers the maximal SNR of the
standard-custodial cloud from $\simeq 141$ (\texttt{BSMPT}) to $\simeq 56$
(\texttt{TL}), but does not move any point across the detection threshold once HL-LHC
testability is imposed, such that the testable fraction of the LISA-observable points agrees between
the two codes to within one percentage point.

The two pipelines differ not only
in their treatment of the transition: \texttt{BSMPT} assumes an observation time of $3\,$yr and a
LISA noise model without an astrophysical foreground, whereas \texttt{TL} assumes
$4\,$yr and includes the confusion noise from unresolved galactic binaries
(section~\ref{subsec:SNR_LISA}). The comparison is thus not a unit test of two implementations of one
prescription, but a comparison across the span of currently defensible choices, from the equation of
state and the soundwave lifetime down to the observation time and the noise budget. That the
HL-LHC-testable fraction of the LISA-observable points agrees to within one percentage point across
all of them is why we regard the complementarity result as robust rather than as a property of one
particular pipeline.

The one modelling choice that we have tested beyond such a code comparison is the bubble wall
velocity, which sets both the kinetic-energy efficiency $\kappa_\text{sw}(v_\text{w},\alpha)$ and the
mean bubble separation $R H_*$. Since neither of the two LTE prescriptions used in \texttt{TL}
and \texttt{BSMPT} can describe a
wall that is slowed down by out-of-equilibrium (OOE) friction, we have recomputed $v_\text{w}$ with
\texttt{WallGo}~\cite{Ekstedt:2024fyq} for a selection of 100 standard-custodial points, distributed
uniformly in
$(\alpha,\beta/H_*)$ and drawn in equal parts from the LISA-observable and from the borderline ($1 <
\text{SNR}_\text{LISA} < 10$) subsets, treating six OOE species ($t_\text{L,R}$, $W$,
$H^\pm_\text{R,I}$, $A$) as recommended in ref.~\cite{vandeVis:2025plm}. 
None of the points for which the solvers converge yields a sub-luminal steady-state wall, as one expects for strong and fast transitions in a
model without a light vector portal, where the LO friction only grows logarithmically with
$\gamma_\text{w}$.\footnote{For $56$ of the $100$ points both the leading-order deflagration and
detonation solvers converge and return a negative net wall pressure at every $v_\text{w} \in (0,1)$,
i.e.~these are genuine runaways at this order. For $43$ points the Boltzmann solver instead aborts
with a singular kinetic matrix, on both branches; these have an only $\simeq 40\,\%$ larger median
$(\beta/H)_{RH}$ than the converged ones and are therefore expected to be at least as relativistic.
The remaining point never reached \texttt{WallGo}: it is a perfectly regular point of the chain,
for which the main analysis returns $\alpha = 0.13$ and $\text{SNR} = 2.1$, but the interface to
\texttt{WallGo} re-traces the phases from scratch in order to hand them over at $T_\text{perc}$, and
it is this second, independent tracing which did not reproduce the zero-temperature minimum. We
further patched a tuple-unpacking bug in the detonation solver of \texttt{WallGo} to obtain these
results and reported it to its maintainers, who have fixed it upstream; the fix is contained in
\texttt{WallGo v1.1.2}, whereas the hybrid-regime singularity remains open at the time of writing.} This confirms the
$v_\text{w} = 1$ fallback of the main analysis performed in \texttt{TL} up to the NLO
transition-radiation terms of
ref.~\cite{Gouttenoire:2021kjv}, and it in particular leaves no borderline point lifted above the
detection threshold, which would have been the only way for the wall-velocity treatment to reduce the
HL-LHC-testable fraction. We have not repeated this exercise for the twisted-custodial cloud, whose
smaller $\alpha$ makes a finite OOE wall velocity more likely, but which is already testable at the
HL-LHC in essentially its entirety.

We hence conclude that the made statistical claim is robust under the theoretical uncertainties
discussed here. The
modelling choices discussed above amount to an $\mathcal{O}(1)$ uncertainty on the SNR of an
individual model realization, but leave the complementarity result of section~\ref{sec:results}
unchanged.

\section{Correlation of 2HDM model parameters with strong transitions}
\label{app:corner}
\begin{figure}[htb]
    \centering
    \includegraphics[width=\textwidth]{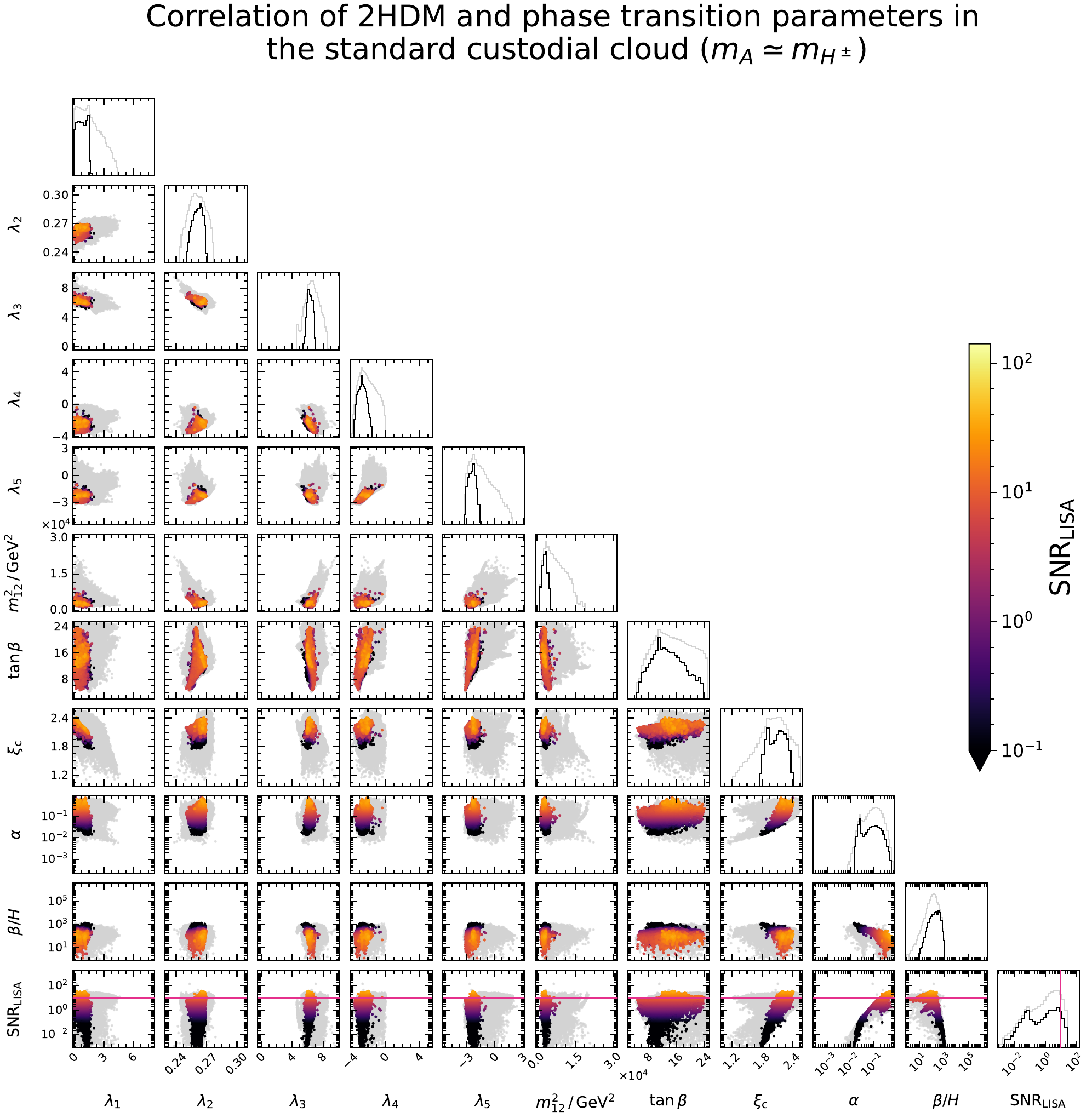}
    \caption{Correlation between the free 2HDM parameters in the coupling basis and GW
    parameters in the standard-custodial cloud. Each off-diagonal panel shows a scatter
    plot of two parameters. Points featuring high-temperature EWSR are colour-coded by
    $\log_{10}(\mathrm{SNR}{_\mathrm{LISA}})$, with points below $\mathrm{SNR}{_\mathrm{LISA}}<0.1$
    shown in black and non-EWSR points are shown in light grey. The diagonal panels show the
    corresponding distributions as binned point counts, with EWSR and non-EWSR points shown in
    black and light grey, respectively. The pink line in the lower row of panels, and in the
    rightmost panel of the diagonal, indicates the LISA detectability threshold,
    $\mathrm{SNR}{_\mathrm{LISA}}=10$.
    }
    \label{fig:cornerA_coupling}
\end{figure}

\begin{figure}[htb]
    \centering
    \includegraphics[width=\textwidth]{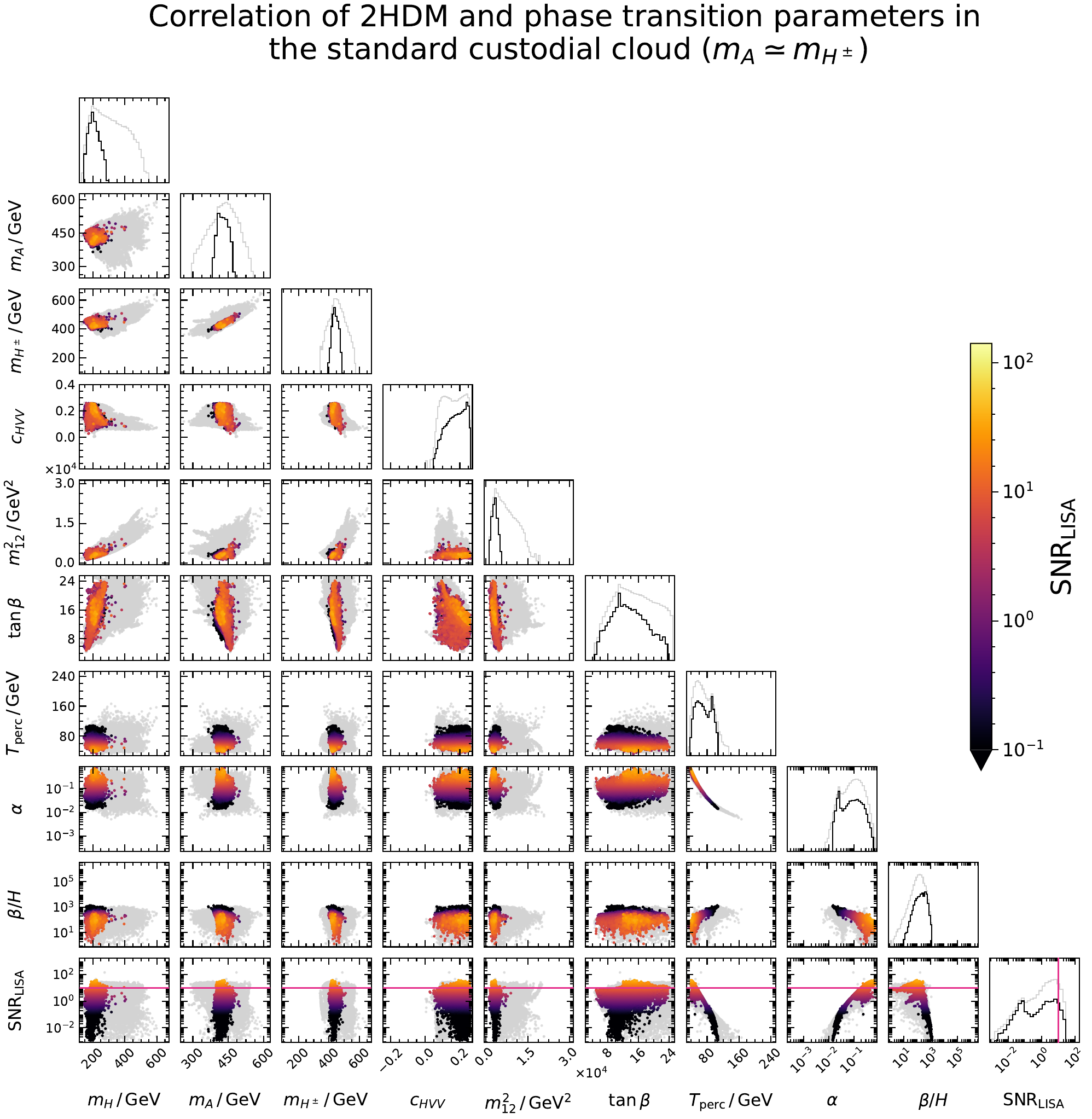}
    \caption{Same as figure~\ref{fig:cornerA_coupling} in the mass basis, with the percolation
    temperature shown in place of $\xi_\text{c}$. The strong-first-order preselection
    $\xi_\text{c}\ge1$ is still applied; the ${\sim}1.5\,\%$ of points for which \texttt{BSMPT}
    returns no percolation temperature are absent from these two panels but retained everywhere else.}
    \label{fig:cornerA_mass}
\end{figure}

\begin{figure}[htb]
    \centering
    \includegraphics[width=\textwidth]{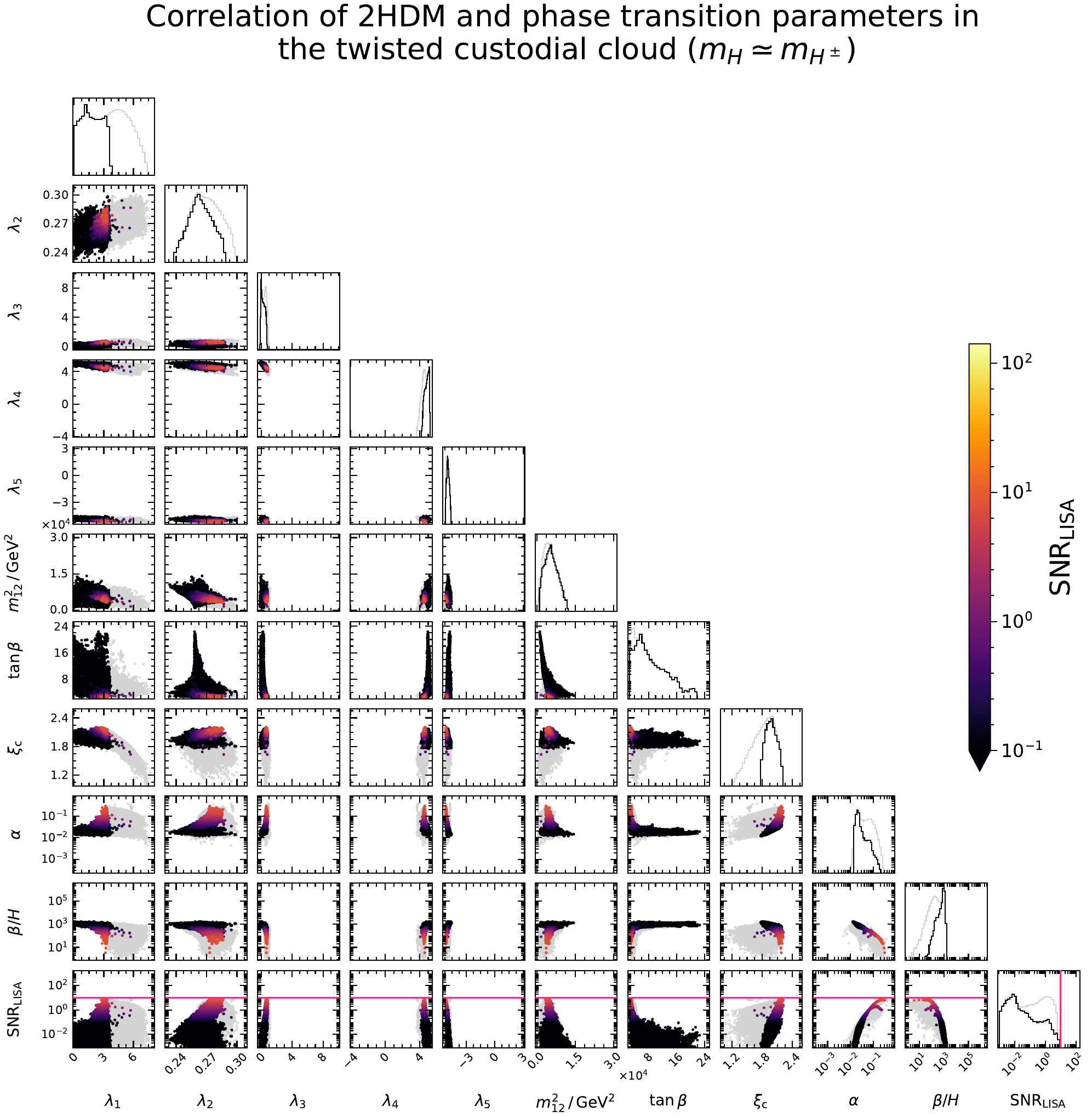}
    \caption{Same as figure~\ref{fig:cornerA_coupling} for the twisted-custodial cloud.}
    \label{fig:cornerB_coupling}
\end{figure}

\begin{figure}[htb]
    \centering
    \includegraphics[width=\textwidth]{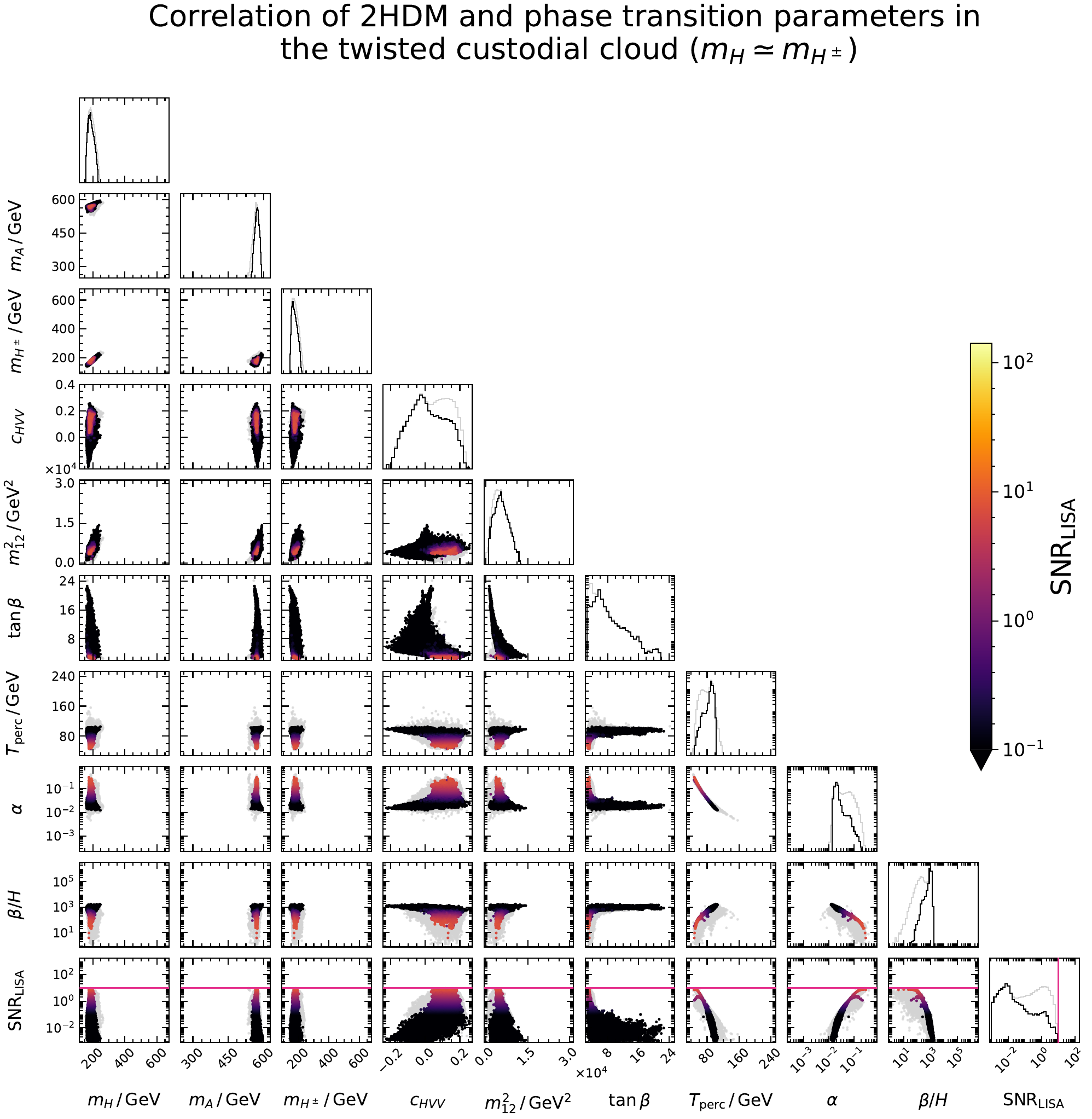}
    \caption{Same as figure~\ref{fig:cornerA_mass} for the twisted-custodial cloud in the mass basis.}
    \label{fig:cornerB_mass}
\end{figure}

In figures~\ref{fig:cornerA_coupling} to~\ref{fig:cornerB_mass}, the correlation between the
2HDM model parameters in the coupling and mass basis and GW parameters are displayed.
Figures~\ref{fig:cornerA_coupling} and~\ref{fig:cornerA_mass} display the standard-custodial
cloud ($m_A\simeq m_{H^\pm}$) in the coupling and in the mass basis, and
figures~\ref{fig:cornerB_coupling} and~\ref{fig:cornerB_mass} show the twisted-custodial
cloud ($m_H\simeq m_{H^\pm}$) in the same two bases. In each corner plot, colour-coded
points refer to high-$T$ electroweak symmetry-restoring model realizations and show the
respective SNR at LISA, whereas grey points do not feature EWSR at $T \to \infty$.

All quantities shown in this appendix are \texttt{BSMPT} predictions; the corresponding
\texttt{TL} results enter the point-by-point comparison of
appendix~\ref{app:uncertainties}. The axis ranges and the colour scale are deliberately common to
all four figures, so that the two clouds can be read against each other directly. All four
figures show every point passing the strong first-order preselection $\xi_\text{c}\ge1$ with
$\mathrm{SNR}>10^{-3}$, split by whether electroweak symmetry is restored; the ranges and medians
quoted below therefore refer to that population, which is considerably broader than the
$\mathrm{SNR}>1$ and $\mathrm{SNR}>10$ subsets characterized in section~\ref{sec:results}.

In the coupling basis of figure~\ref{fig:cornerA_coupling}, the LISA-observable region of the
standard-custodial cloud occupies a compact but not fine-tuned part of the parameter space:
$\lambda_1\lesssim3$, $\lambda_2$ sharply centred on $0.26$, $\lambda_3\simeq6\text{--}7$ and
$\lambda_4\simeq\lambda_5\simeq-2$ to $-3$, with $m_{12}^2\simeq2\text{--}12\times10^3\,\text{GeV}^2$ and
$\tan\beta\simeq8\text{--}23$. The SNR colour coding shows clear gradients along essentially every
direction, so the loudest signals are reached well inside the sampled region rather than at its
boundary; the scan is not pushing against one of its own cuts. Large $\alpha$ maps directly onto
large SNR, and the two measures of transition strength are themselves tightly correlated, reaching
$\alpha\simeq0.7$ and $\xi_\text{c}\simeq2.6$. The inverse duration peaks at a few hundred, consistent with the
$10\text{--}15\,\%$ larger $(\beta/H)_{RH}$ that \texttt{TL} returns for the same points
(appendix~\ref{app:uncertainties}).
Dropping the requirement of electroweak symmetry restoration opens the parameter space in almost
every direction and admits a larger fraction of high-SNR points, but it does not do so uniformly:
several coupling combinations stay closed, held shut by the theoretical and collider constraints of
section~\ref{sec:paramspace} rather than by the phase transition.

Figure~\ref{fig:cornerA_mass} shows the same points in the mass basis. The custodial relation is
explicit, $m_A\simeq m_{H^\pm}$ at $400\text{--}460\,$GeV, while $m_H$ stays light at $165\text{--}250\,$GeV in the
restoring subset. The vector coupling reaches $c_{HVV}\simeq0.25$, and within the EWSR subset the
largest SNRs sit at the largest $c_{HVV}$. This is what makes the subset easy to test, since the
same coupling drives $gg\to H\to ZZ$. Without EWSR, smaller $c_{HVV}$ are also
allowed, and it is this tail that reduces the collider coverage. The
percolation temperature spans $44\text{--}99\,$GeV and is very nearly a proxy for the transition strength
on its own: across the cloud its Spearman rank correlation with $\log\alpha$ is $-0.99$, and with
$\log\mathrm{SNR}$ it is $-0.98$. The $T_\text{perc}$ panel is therefore the $\alpha$ panel read
backwards, and it displays directly the mechanism identified in section~\ref{subsec:cloudB}, namely
that what sets the transition strength here is the temperature at which the energy is released
rather than the amount released.

The twisted-custodial cloud in figure~\ref{fig:cornerB_coupling} reaches markedly smaller SNRs,
barely touching the detection threshold, and requires considerably more tuning to do so. Its
couplings are confined to narrow bands, $\lambda_3\simeq-0.04$ to $0.6$, $\lambda_4\simeq4.5\text{--}5.2$
and $\lambda_5\simeq-5.1$ to $-4.8$, i.e.~$\lambda_4\simeq-\lambda_5$, the twisted relation of
appendix~\ref{app:custodial}. The maximal $\xi_\text{c}\simeq2.2$ stays below the
$2.6$ of the standard-custodial cloud. Most striking is the anticorrelation between $\alpha$ and
$\tan\beta$: the strongest transitions occur only at the lowest $\tan\beta$, the top five per cent
in $\alpha$ having $\tan\beta\simeq2.5\text{--}4$ against a full range of $2.5\text{--}22$. This is the mechanism of
section~\ref{subsec:cloudB} operating within a single cloud rather than between the two: a low
$\tan\beta$ lowers the percolation temperature, and the transition strength follows through
$\alpha\propto T_\text{perc}^{-4}$. Relaxing EWSR reproduces
the same pattern with only a mildly enlarged parameter space, the exception being the $\lambda_1$
direction, where the restoring subset reaches $\lambda_1\lesssim3.3$ and the non-restoring one
extends to $6\text{--}8$.

In the mass basis, figure~\ref{fig:cornerB_mass}, the twisted degeneracy $m_H\simeq m_{H^\pm}$
is realized between $165$ and $210\,$GeV, with a heavy pseudoscalar at
$m_A\simeq555\text{--}585\,$GeV. That mass lies
well above $2m_t$, so $A\to t\bar t$ is open throughout this cloud --- the channel discussed and set
aside in section~\ref{subsec:caveats}. The strongest transitions, with and without EWSR, are found
for vector couplings $c_{HVV}\simeq0.05\text{--}0.15$. The larger SNRs of the non-restoring subset come
almost entirely from its smaller inverse duration, the median $\beta/H$ falling from $940$ to $190$,
which is in turn permitted by the extra room in the $\tan\beta$ direction opened up along
$\lambda_1$ and $\lambda_2$; the scalar masses themselves are barely affected. The percolation
temperature covers $56\text{--}101\,$GeV, systematically above the $44\text{--}99\,$GeV of the standard-custodial
cloud, which is the reason this cloud is the quiet one for LISA.

\clearpage
\newpage
\bibliographystyle{JHEP_improved}
\bibliography{literature}

\end{document}